\documentclass[]{mn2e}
\usepackage{graphicx}
\usepackage{epstopdf}
\usepackage{verbatim}
\usepackage[T1]{fontenc}
\usepackage{aecompl}

\def\lsun{L$_\odot$}
\def\msun{M$_\odot$}

\def\etal{{\it et al.~}}

\def\lsim{\mathrel{\rlap{\lower4pt\hbox{\hskip1pt$\sim$}}    \raise1pt\hbox{$<$}}} 
\def\aj{AJ}
\def\apj{ApJ}
\def\apjs{ApJS}
\def\apjl{ApJL}
\def\mnras{MNRAS}
\def\aap{A\&A}
\def\araa{ARA\&A}
\def\pasp{PASP}
\def\procspie{Proc. SPIE}

\title[Star clusters in M33] {Evidence for temporal evolution in the M33 disc as traced by its star clusters}

\author[M. A. Beasley \etal]{Michael A. Beasley$^{1, 2}$\thanks{E-mail: \mbox{beasley@iac.es} (MAB); 
\mbox{izaskun@astro-udec.cl} (ISR); \mbox{carme@iac.es} (CG); \mbox{ata@astro.ufl.edu} (AS); 
\mbox{aaj@iac.es} (AAJ)}, Izaskun San Roman$^{3,5}$\footnotemark[1], Carme Gallart$^{1,2}$\footnotemark[1],
\newauthor
 Ata Sarajedini$^{4}$\footnotemark[1], Antonio Aparicio$^{1,2}$\footnotemark[1]\\ 
$^1$ Instituto de Astrof\'\i sica de Canarias. Calle V\'\i a L\'actea s/n. E38200 - La Laguna, Tenerife, 
Canary Islands, Spain.\\
$^2$ University of La Laguna. Avda. Astrof\'isico Fco. S\'anchez, s/n. E38206, La Laguna, Tenerife, 
Canary Islands, Spain.\\
$^3$ Departamento de Astronom\'ia, Universidad de Concepci\'on, Casilla 160-C, Concepcion, Chile\\
$^4$ Department of Astronomy, University of Florida, 211 Bryant Space Science Center
Gainesville, FL 32611, USA\\
$^5$Centro de Estudios de F\'isica del Cosmos de Aragon, Plaza San Juan 1, Planta-2, 44001, Teruel, Spain}

\begin{document}

\date{}

\pagerange{\pageref{firstpage}--\pageref{lastpage}} \pubyear{2014}

\maketitle

\begin{abstract}
 We present precision radial velocities and stellar population
parameters for 77 star clusters in the Local Group galaxy M33.
Our GTC and WHT observations sample both young, massive clusters and known/candidate 
globular clusters, spanning ages $\sim$ $10^{6} - 10^{10}$ yr, and 
metallicities, [M/H] $\sim-1.7$ to solar.
The cluster system exhibits an age-metallicity relation; the youngest clusters 
are the most metal-rich. When compared to HI data, clusters with [M/H]$\sim-1.0$ and younger 
than $\sim4$ Gyr are clearly identified as a disc population.
The clusters show evidence for strong time evolution in the disc radial 
metallicity gradient (d[M/H]dt / dR = $0.03$ dex/kpc/Gyr). 
The oldest clusters have stronger, more negative gradients than the youngest
clusters in M33.  The clusters also show a clear age-velocity dispersion relation. 
The line of sight velocity dispersions of the clusters increases with age similar to Milky Way 
open clusters and stars.  The general shape of the relation is reproduced 
by disc heating simulations, and the similarity between the relations in M33 
and the Milky Way suggests that heating by substructure, and cooling of the ISM both play a  
role in shaping this relation. 
We identify 12 ``classical'' GCs, six of which are newly identified GC candidates. 
The GCs are more metal-rich than Milky Way halo clusters, and show weak rotation. 
The inner ($R<4.5$ kpc) GCs exhibit a steep radial metallicity gradient 
(d[M/H]/dR = $-0.29\pm0.11$ dex/kpc) and an exponential-like surface density profile.  
We argue that these inner GCs are thick disc rather than halo objects.
\end{abstract}
%
\section{Introduction}
\label{sec:Introduction}

Observations suggest that the inner regions of relatively massive disc galaxies 
contain the oldest stars, and that their scalelengths increase with time (e.g. de Jong 1996). 
Such galaxies are thought to complete central star formation early, while their outer regions 
continue to grow as gas at large radii is converted into stars.
Although this phenomenon has been known for two decades, it is now 
generally termed {\it ``inside-out''} growth (e.g., MacArthur et al. 2004; 
Williams et al. 2009; S\'anchez-Bl\'azquez et al. 2014). 
The situation for lower-mass discs  is less clear. 
The Large Magellanic Cloud (LMC; $M_{*}\sim2-3\times 10^{9}$~M$_{\odot}$; Kim et al. 1998; 
van der Marel 2004) - perhaps the best studied example - exhibits young stars that are 
found to be more centrally concentrated than older stars (e.g. Meschin et al. 2014). 
However, this may be in part due to its close proximity to the Milky Way, where 
interaction may act to shrink the star forming H{\rm I} gas disc over time (Nidever 2014). 
The intermediate-mass Triangulum galaxy (M33; $M_{*}\sim3-6\times 10^{9}$~M$_{\odot}$; 
Corbelli 2003) appears to be an admixture of these two mass regimes 
(e.g.,  Barker et al. 2007; Williams et al. 2013; Robles-Valdez et al. 2013). 
The inner disc may form inside-out, and the outer (beyond the truncation radius) disc 
may form ``outside-in''. Due to this dual nature, understanding intermediate-mass 
discs is of key importance for building a comprehensive understanding of 
disc formation (e.g. see  Gogarten et al. 2010 for the case of NGC~300).

From the point of view of stellar population studies, important observational diagnostics 
of disc growth and evolution include the age-metallicity
relation (AMR) of the disc stellar population(s), their radial age gradients
and metallicity gradients (e.g., Freeman \& Bland-Hawthorn 2002; MacArthur et al. 2004). 
For example, one approach to understand disc formation is to 
study the radial metallicity gradients of disc stars and how these gradients evolve with time.
The time evolution of metallicity gradients in discs is important because it is
directly linked to disc growth: gradients encode the location and metallicity of 
gas accreted to the disc. The assumption that stars stay where they were born
implicitly or explicitly form part of many chemical evolution models
of galaxy discs (e.g. Tinsley et. al. 1974; 
Matteucci \& Francois 1989; Boissier \& Prantzos 1999).
However, Ro\v{s}kar et al. (2008) have shown 
that subsequent dynamical evolution of disc stellar populations --
such as radial mixing -- may flatten out such gradients (but see also 
Grand, Kawata \& Cropper 2013). In addition, 
radial mixing tends to flatten and create spread in the disc AMR 
(Ro\v{s}kar et al. 2008; Grand et al. 2014).
The lesson here is that, as emphasised for example 
by Schoenrich \& Binney (2009), disc chemistry and disc kinematics 
cannot be separated if a complete picture of disc formation 
is to be obtained (a good example of this approach is the CALIFA survey - 
see e.g., S\'anchez-Bl\'azquez et al. 2014).
 
In terms of kinematics, an important time-dependent diagnostic of disc
formation is the age-velocity dispersion relation. The velocity
dispersion of disc stars in the Milky Way increases with age (e.g. Holmberg et al. 2007;
Soubiran et al. 2008; Aumer \& Binney 2009) and recent work suggest 
this is also the case in M31 (Dorman et al. 2015).
The origin of these relations is still unclear, but might arise due to either disc heating 
processes (e.g.  Spitzer \& Schwarzschild 1951; Wielen 1977; Mihos \& Hernquist 1996; 
Minchev \& Quillen 2006;  H\"anninen \& Flynn 2002; Martig, Minchev \& Flynn 2014), 
or the different kinematical properties of gas at the time of star formation 
(e.g. Brook et al. 2004; Bournard, Elmegreen \& Martig 2009; Bird et al. 2013), 
or some combination of the two mechanisms.

Observationally, one way of studying the time evolution of disc populations is to go to high 
redshift.  For example, Yuan et al. (2011) studied the HII regions of a lensed disc 
galaxy at $z=1.49$ and found a negative metallicity gradient more than ten times 
steeper than nearby systems - suggestive that metallicity gradients may flatten over time.
However, major difficulties in this approach are that at $z>1$ discs 
subtend only a few arcseconds. The information obtained from such studies is 
limited by the small spatial scales and faintness of these systems.

An alternative approach is to study  nearby discs in detail, and compare 
the properties obtained from stellar populations that trace different epochs 
of star formation.  Chemical abundances for massive (OB) stars are feasible for the nearest galaxies 
(e.g. Urbaneja et al. 2005; Esteban et al. 2009), while studies of more 
evolved populations have concentrated on bright emission lines 
from Planetary Nebula (e.g. Magrini et al. 2004), or on RGB stars for the very nearest 
systems (e.g., Kalirai et al. 2006; Leaman et al. 2009; Carrera et al 2011).
While extremely useful, these tracer populations generally represent stellar populations with a limited 
range of (or poorly constrained) ages. For example, OB stars have lifetimes of $\sim$Myr and therefore 
reflect very recent star formation. In contrast PNe and RGB stars can span a wide range of ages, 
but their precise ages are hard to constrain - particularly for the oldest populations.

In contrast, star clusters in late-type galaxies 
exhibit a wide range of ages, from very young, massive clusters up to GC 
ages, present the opportunity to study disc evolution and kinematics over 
the whole lifetime of the galaxy (e.g., Huchra, Brodie \& Kent 1991; Chandar et al. 2002; 
Schroder et al. 2002; Beasley et al. 2004; Bridges et al. 2007; Caldwell et al. 2011).
In addition, star clusters can, in principle, have their ages and metallicities determined
via studies of their resolved stars,  or through integrated light techniques.

Here we present the first results of a spectroscopic campaign to characterise
the star cluster system of M33. Due to its mass and proximity, M33 presents an important
laboratory for understanding the formation of disc galaxies, and addressing some
of the above issues. M33 is an intermediate mass 
($M_{*}\sim3-6\times 10^{9}$~M$_{\odot}$; Corbelli 2003) disc system 
with a minimal bulge component (e.g. Regan \& Vogel 1994).  
Its proximity ($847\pm60$~kpc; Galleti, Bellazzini \& Ferraro 2004) and favourable inclination 
angle ($i=56^{\circ}$; Paturel et al. 2003) make the galaxy an ideal target for 
understanding the disc ISM and its stellar populations.
Star formation histories have been obtained for M33 based on both small-area, deep {\it HST}
photometry (e.g. Barker et al. 2007; Williams et al. 2009) and on wider-field, 
but shallower studies (e.g., Davidge \& Puzia 2011).

The star cluster system of M33 has been previously studied by a number
of authors (see Hodge 2012 for a review). Notable imaging campaigns include those
of Hiltner (1960), Christian \& Schommer (1982), Chandar, Bianchi \& Ford 1999, 
Sarajedini \& Mancone (2007), Park \& Lee (2007) and San Roman, Sarajedini \& Aparicio (2010). 
In total, some $\sim600$ cluster candidates have been 
identified\footnote{http://www.mancone.net/m33\_catalog/}.
Spectroscopically, the M33 star clusters have been studied by Schommer et al. (1991), 
Chandar et al. (2002; hereafter CBFS02), Chandar et al. (2006) and  
Sharina et al. (2010). CBFS02 presented the most detailed kinematic 
study to date. They obtained velocities and estimated ages from a combination of 
integrated spectroscopy and integrated colours for 107 M33 star clusters and 
were able to separate the cluster system into 
two populations: a young population with disc kinematics and an older population
with kinematics consistent with a halo cluster system. They also argued
for an age spread of $\sim6$ Gyr amongst the halo population. 
However, CBFS02 did not derive metallicities for their cluster sample
and most ages came from integrated colours which are strongly affected
by age-metallicity degeneracy.

In the following, we use the term {\it star clusters} to refer
to stellar clusters as a general class (open, massive and globular), referring 
to globular clusters in the classical sense as the analogues of the Milky Way 
GCs. We assume a distance of 847 kpc to M33 (Galleti et al. 2004) 
for all conversions to physical distances and an optical disc scale
of 9.2 arcminutes (Guidoni et al. 1981) which corresponds to 2.3 kpc at our
adopted distance.
In Section~\ref{TheData} we describe our observations and data reduction
procedure. In Section~\ref{Analysis} we show how cluster radial 
velocities and stellar population properties are derived.
In Section~\ref{Results} we discuss our results for the 
stellar population properties and kinematics of the 
M33 star clusters. We summarise and discuss our findings in Section~\ref{Conclusions}. 

%

\section{The Data}
\label{TheData}

\subsection{The cluster sample and methodology}
\label{subsec:methodology}

Our observing strategy consisted of using the WYFFOS/AF2 multifibre spectrograph 
(Watson 1995; Bridges 1998) on the William Herschel Telescope (WHT) to focus on the 
brightest M33 clusters, and OSIRIS (Cepa et al. 2000) in longslit mode on the 
Gran Telescopio de Canarias (GTC) to cover the fainter end of the luminosity function. 
Our star cluster sample was chosen from the updated version catalogue of Sarajedini \& Mancone (2007)
in addition to the newer catalogue of San Roman et al. (2010). 
The clusters were selected in order to cover as much of the optical disc 
as possible. All the targets posses accurate positions and integrated 
magnitudes based on {\it HST} and/or ground-based observations. 

The GTC sample consists of confirmed clusters (from high 
resolution imaging or spectroscopy) with $g\prime > 19$. The WHT observations 
prioritize the observations of bright confirmed clusters ($g\prime < 19$), but to 
maximize the number of fibres used in each exposure, these requirements 
were relaxed and extra fibres were allocated for less certain, fainter 
candidate clusters. This potentially introduced some contamination into the sample 
(see Section~\ref{RadialVelocities}).  

We also included in the sample several objects to be used as radial velocity templates 
(but see Section~\ref{RadialVelocities}). 
The M33 nucleus was observed as a template for young clusters while known 
GCs R12=SR1710=SM316 and 
R14=SR1765=SM275 (Sarajedini et al. 1998; Larsen et al. 2002) were observed as
more appropriate templates for older objects. In addition, clusters  M9=SR2075=SM402, 
U49=SR1458=SM178 and H38=SR1566=SM206 (from the same studies) were included as 
scientifically interesting cases based on their similarities to 
Milky Way GCs. 

\subsubsection{GTC data}

Previously confirmed candidates (identified by radial velocities 
or high resolution imaging) were  observed using OSIRIS on the 
Gran Telescopio de Canarias (GTC) in long-slit mode during
semesters 2010B$-$2012A.
A R2500V grism was used yielding an effective wavelength range of 4500-5600~\AA.
A slit-width of 1.0 arcsec was used in all cases yielding a 
final spectral resolution (FWHM) of 2.1~\AA. 
In general, two objects were placed on each slit- with the exception of 
one observing block where 3 objects were placed on the slit. 
Integration times were 900$-$5400s per object (depending upon the target magnitude)
and we observed a total of 53 unique star clusters and obtained useable spectra for 48.  
A summary of the GTC observations is given in Table~\ref{tab:setup}.

Basic reductions were performed using the OOPS OSIRIS data reduction pipeline 
written in PyRAF. These data were debiased, flat-fielded and wavelength solutions
obtained using 20 to 30 Xenon, Neon and Argon arclines. 
Typical residuals to the wavelength solution 
were 0.08~\AA. For some spectra we noticed that the 5577.34~\AA\ skyline was shifted by up to 
0.2~\AA\ from the rest frame value, and in these cases we shifted the spectra in 
wavelength to the correct value. The final extraction of the 1-d spectra was performed using 
APALL in pyRAF. The spectra were flux-calibrated using the standards L1363-3 and 
BD28+4211 and with tasks STANDARD, SENSFUNC and RESPONSE also in pyIRAF. 
The final, extracted spectra have a median S/N of 40 per~\AA.

\subsubsection{WHT data}

Previously identified star clusters, and star cluster candidates identified by 
San Roman et al. (2010) were observed using the WYFFOS/AF2 multifibre instrument 
on the William Herschel Telescope in La Palma.  
WYFFOS/AF2 allows for the allocation of up to 160 fibres over a 40 arcminute field of view. 
Two fibre configurations were observed over a total of five nights of observations, one 
configuration during the nights of 7$-$8 October and
8 November 2010, the second configuration during the nights of 23 and 24 October 2011. 
We used the 1200B grism which gave a useful spectral range of 
4000--5500~\AA\ and a spectral resolution of 2.1~\AA\ (FWHM).
A summary of the WHT observations is given in Table~\ref{tab:setup}.

In order to maximise the number of clusters observed with the WHT, and taking into 
account fibre positioning limitations, we used two different fibre configurations. 
Half of the fibres were allocated in both configurations to observe the faintest 
objects of the WHT sample; the remaining fibres were used for the brightest 
objects. Each WYFFOS/AF2 fibre has 1.6 arcsecond diameter, similar to the size of the star 
clusters ($0.8 < r < 2.0$ arcseconds), thereby minimizing any loss of light. Few of these 
objects have previous velocity measurements.

Because the M33 disc provides a bright and variable background emission, 
estimations of the underlying galaxy light should be performed as locally as possible. 
For this reason, the observations were taken in a "beam-switching" mode, with 
three 15 minute off-source integrations (offset from the objects by 10 arcseconds to the 
north, west and east) interspersed within the 30 minute integrations on the clusters. 
Dedicated sky fibers were also placed across the field to check the background level. 
Comparison of the sky residuals indicated that the offset sky fibres 
gave better results (smaller sky residuals) than the dedicated sky fibres
and therefore we used these to remove the sky contribution.
The rms difference in the sky level across the sky fibres, after removing those
fibres that fell on bright sources in the offset skies, was typically 10-15 percent.
For the brightest targets, the sky background comprised $\sim5$ percent of the total
light down the fibre, rising to $\sim25$ percent for the faintest objects.
These data were reduced using the dedicated IDL WYFFOS/AF2 reduction pipeline (v1.02).
The software performs bias subtraction, flat-field corrections using sky-flats
and allows for the determination of the wavelength solution. For this we 
used Helium and Neon arc lines which gave typical residuals of 0.1~\AA. 
Finally, a relative flux calibration was obtained using the standard BD+28 4211.

The final, extracted spectra have a median S/N of 20 per~\AA.
A number of the spectra have significantly lower (typically a factor of 2 lower) 
S/N than expected based on the WYFFOS/AF2 ITC. 
We attribute this to problems with fibre positioning known
to affect the instrument prior to the 2013 upgrade. After merging overlapping 
objects in the WHT samples, we obtained a total of 61 spectra for which we could
measure radial velocities of which 29 spectra have sufficient S/N for 
stellar population analysis.

\begin{table}
\begin{center}
\centering
\caption[Instrumental Setup]
{The instrumental setups for spectroscopy.}
\begin{tabular}{ll} 
\hline 
Telescope & GTC 10.4 m\\
Instrument & OSIRIS longslit mode (1$\farcs$ slit)\\
Dates & queue mode semesters 2010B$-$2012A\\
VPH grating & 2500V\\
Spectral range & 4500--6000 \AA \\
Dispersion & 0.80 \AA~pixel$^{-1}$\\
Resolution (FWHM) & $\sim$ 2.1 \AA \\
Detector & 2 Marconi CCD42-82 (2048 $\times$ 4096 pixels) \\
Gain & 0.95 e$^{-}$~ADU$^{-1}$\\
Readout noise & $\sim$4.5 e$^{-}$ \\
Seeing & 0.8--1.2$\farcs$ \\
\hline
\end{tabular}
\begin{tabular}{ll} 
\hline 
Telescope & WHT 4.2 m\\
Instrument & AF2 multifibre spectrograph (1.6$\farcs$ fibres)\\
Dates & $7-8$ Oct., 8 Nov. 2010, $23-24$ Oct. 2011 \\
Grating & 1200B\\
Spectral range & 4575--6000 \AA \\
Dispersion & 0.43 \AA~pixel$^{-1}$\\
Resolution (FWHM) & $\sim$ 2.1 \AA \\
Detector & 2 EEV-42-80 (4096 $\times$ 4096 pixels) \\
Gain & 0.90 e$^{-}$~ADU$^{-1}$\\
Readout noise & $\sim$4.2 e$^{-}$ \\
Seeing & 0.9--1.4$\farcs$ \\
\hline
\end{tabular}
\label{tab:setup}
\end{center}
\end{table}

\section{Analysis}
\label{Analysis}

\subsection{Radial Velocities}
\label{RadialVelocities}

Radial velocities for the star clusters and cluster candidates were 
determined by fourier cross-correlation
using FXCOR in pyIRAF. For cross-correlation templates, we used a total of 350 MILES 
model SEDs (Vazdekis et al. 2010) in the age range 63 Myr to 17 Gyr and metallicity 
([M/H]) range $-2.3$ to +0.2.
The resolution of our data ($\sim2.1$~\AA\ FWHM) is slightly higher than that of MILES 
(2.5~\AA\ FWHM) and so we rebinned our spectra in order to match resolutions for 
cross-correlation.  
Despite the mild degradation in spectral resolution, tests showed that this turned out 
to be a better approach than using the template stars we observed at the native WHT and 
GTC resolutions. Even though, in principle, the slightly higher native spectral 
resolution should give higher velocity precision, in practice a lack of a wide range 
of adequate templates can prevent this precision from being achieved. This template 
mismatch problem is particularly acute for a cluster system such as M33 that has a 
very wide range of ages and metallicities (see Section~\ref{StellarPopulations}). 
The best-fitting MILES templates provided significantly better (higher) normalised 
cross-correlation peaks than the best-fitting observed template stars. 
Therefore for each cluster
we took the velocity returned for the best-fitting MILES SED as our final velocity.
Heliocentric corrections were then applied to these velocities.

\begin{figure}
\centerline{\includegraphics[width=9.0cm]{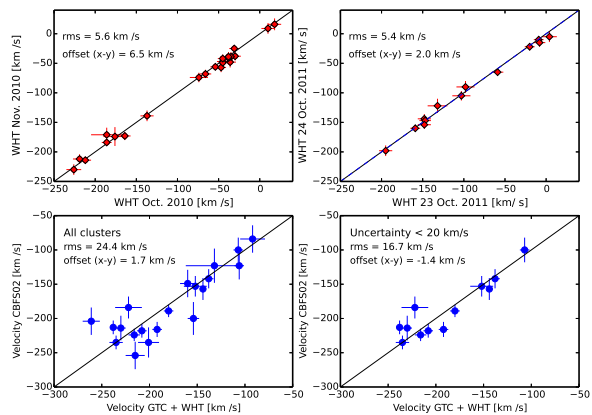} }
\caption{Comparison of cluster velocities in common between different 
observing runs on the WHT ({\it top panels}) and for clusters 
in common with CBFS02 ({\it bottom panels}). Solid lines have unit slope
in all cases. 
\label{cfWHT1}}
\end{figure}

We assessed both the precision and accuracy of our velocities via run-to-run comparisons
and through comparisons with literature values (Figure~\ref{cfWHT1}). 
For velocities with FXCOR Tonry \& Davis (1979)
$R$ values of $R\geq15$ the agreement is good between different WHT runs with mean 
differences of 6.5 km/s (2.0 km/s) and rms 5.6 km/s (5.4 km/s) for the 
October/November 2010 (October 2011) observations and we take $R=15$ 
(corresponding approximately to S/N$\sim10$ per~\AA in these data) 
as the lower limit for reliable velocities for these data. For the GTC observations there are 
unfortunately few (4) overlapping velocities with the WHT sample with which to check our
velocities. However, we find 16 GTC velocities and 4 WHT velocities in common 
with those obtained by CBFS02. Comparisons between the combined
GTC+WHT data with those of CBFS02 are shown in the bottom panels of Figure~\ref{cfWHT1}.
Comparing all the clusters in common we find an offset of us - CBFS02 of +1.7 km/s and
an rms of 24.4 km/s. Selecting CBFS02 clusters with velocity uncertainties 
less than 20 km/s we obtain us - CBFS02 = $-$1.4 km/s with rms 16.7 km/s. 
We also observed the nucleus of M33 with the same observational set up as for the GCs, 
and find heliocentric radial velocities of $-182\pm5$ km/s (GTC) and $-185\pm4$ km/s (WHT), 
in good agreement with the NED value ($-179\pm3$ km/s).

Based on these comparisons we conclude that there is good velocity consistency
both within our dataset, and between our dataset and that of CBFS02. The mean velocity 
uncertainty of our dataset is 9.0 km/s with a standard deviation of 5.4 km/s.

\subsection{Star Cluster Identification}
\label{StarClusterIdentification}

All the GTC targets are genuine star clusters in M33, identified either in 
high-resolution ground-based imaging, from HST observations, or spectroscopically. 
The WHT targets contain a mix of known and candidate star clusters and in these 
data we expected some level of contamination from foreground stars.
Clusters in the M33 disc are expected to follow the HI rotation
(e.g., Schommer et al. 1992; CBFS02) and therefore are kinematically 
separable from Milky Way halo stars based on their well-defined 
velocity distribution and similarity to the disc rotation solution. 
However, any clusters in M33 belonging to kinematically hotter populations 
(e.g., thick disc, halo) are expected to have a broader distribution
of radial velocities.

\begin{figure}
\includegraphics[width=10.0cm]{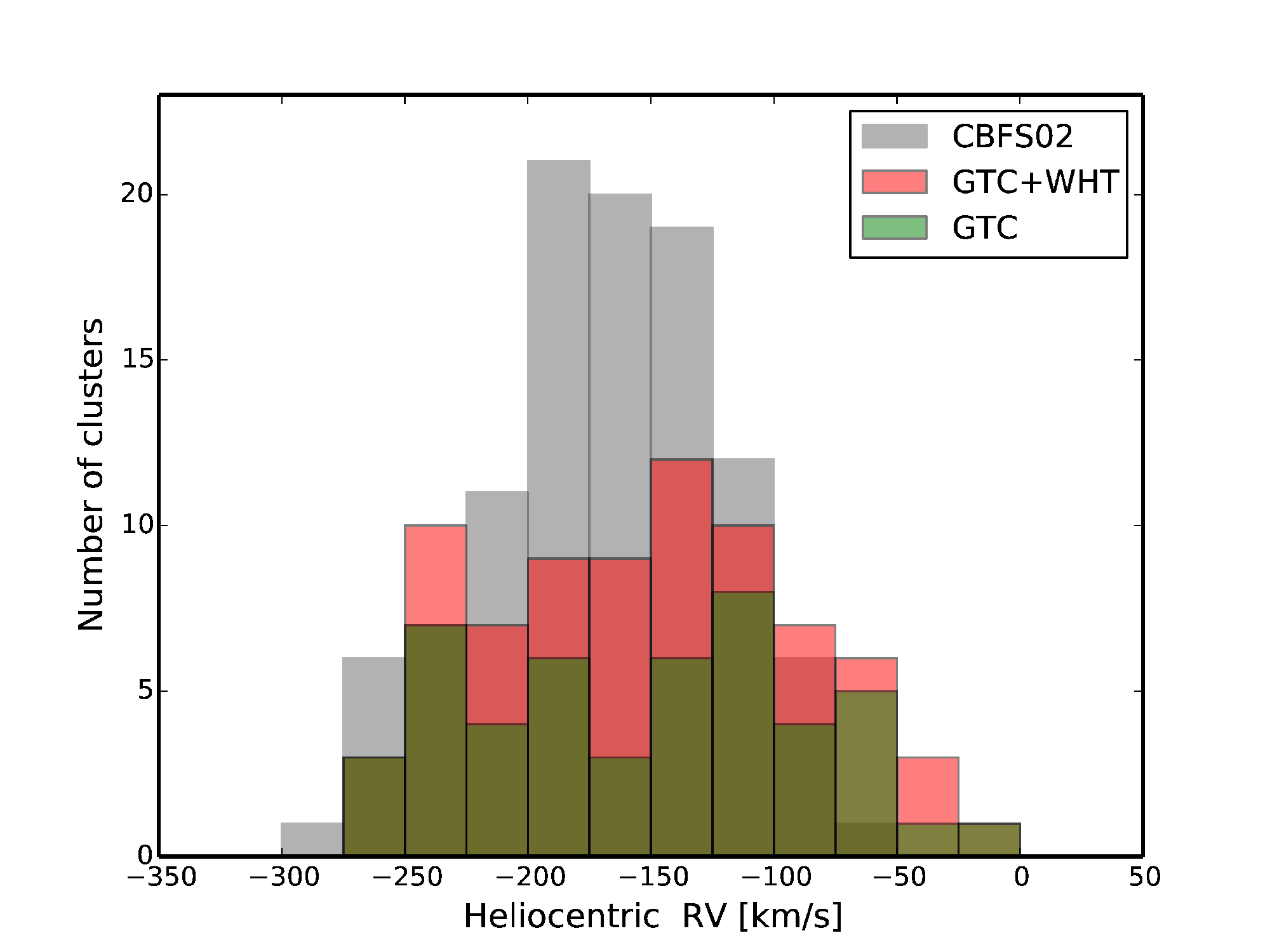} 
\caption{Distributions of the velocities of our GTC and WHT samples
compared to the sample of CBFS02.  The GTC sample shows a wide range
of velocities consistent with a disc distribution.
\label{vel_hist}}
\end{figure}

Therefore, we visually inspected all the candidate WHT spectra to look for 
interloping foreground stars. A number of objects were unambiguously identified 
as foreground stars based on either the presence of very strong 
molecular features (e.g., CN, CH, TiO), or on atomic lines (e.g. NaD) much 
stronger than expected for star clusters.
All these objects have $v_{\rm helio}>$ $-100$ km/s, and we classify them as 
stars and do not consider them further in this analysis.
Five cluster candidates (WHT-72, WHT-84, WHT-286, WHT-318 and WHT-402) 
have spectra resembling old GCs and velocities which are 
consistent with kinematically hot M33 stellar population or low velocity foreground stars. 
Based on the above and their 
non-stellar appearance on ground based imaging we tentatively 
classify these as GCs. An additional candidate (WHT-323) 
has $v_{\rm helio}$ = $-160\pm4$ and we also consider this a good GC candidate. 

The velocity histograms of the GTC and cleaned WHT samples are 
shown in Figure~\ref{vel_hist}. 
For comparison,  we show the cluster velocities from the sample of CBFS02.
The GTC+WHT velocities occupy a velocity range similar to that of CBFS02, 
but it is possible that we are missing some of the lowest velocity 
clusters. Whether this is real or a result of our candidate selection 
is unclear. The GTC velocities show a wide range consistent with a 
disc population.

In total, we identify 77 clusters suitable for stellar population analysis 
and their basic properties are listed in Table~\ref{tab_sample}.

\begin{table*}
\caption{Basic and derived data for M33 cluster sample. Full table available online. IDs are from
San Roman et al. (2010).}
\begin{tabular}{lcccccccl}
\hline \hline
ID & RA (J2000) & Dec (J2000) & Age & [M/H] & V$_{\rm helio}$ & V$_{\rm HI}$ & PA & notes\\  
 & (hh mm ss) & (dd mm ss) & (Gyr) & (dex) & (km/s) & (km/s) & (degrees) &  \\
\hline
856 & 1 32 55.39 & +30 38 38.57 & $3.845\pm0.924$ & $-1.21\pm0.19$ & $-130\pm12$ & $-148$ & 266.0 & \\
902 & 1 32 59.90 & +30 27 19.92 & $0.281\pm0.105$ & $-0.03\pm0.25$ & $-62\pm24$ & $-108$ & 226.1 & \\
941 & 1 33 6.88 & +30 41 3.46 & $1.165\pm0.457$ & $-0.43\pm0.26$ & $-171\pm11$ & $-164$ & 277.5 & \\
1066 & 1 33 16.06 & +30 20 56.66 & $0.19\pm0.032$ & $-0.1\pm0.17$ & $-70\pm7$ & $-101$ & 205.1 & \\
1115 & 1 33 20.53 & +30 49 1.52 & $9.967\pm1.528$ & $-1.13\pm0.17$ & $-160\pm4$ & $-221$ & 321.0 & GC candidate\\
1133 & 1 33 22.13 & +30 40 25.80 & $0.019\pm0.003$ & $-0.16\pm0.16$ & $-160\pm8$ & $-165$ & 276.5 & young, HeI \\
\hline
\end{tabular}
\label{tab_sample}
\end{table*}

\subsection{Stellar population analysis}
\label{StellarPopulations}

To estimate the ages and metallicities of the star 
clusters we primarily used the Vazdekis et al. (2010) stellar population models 
based on the MILES empirical stellar library of S\'anchez Bl\'azquez (2006)
(hereafter the MILES models\footnote{http://miles.iac.es}). 
The youngest age predicted by the MILES  models (version 9.0), 
based on the Padova isochrones (Girardi et al. 2000), is 63~Myr. 
We expected clusters younger than this in our sample and in order to model 
these younger clusters we extended the MILES parameter space with the models 
of Gonz\'alez Delgado (2005; hereafter GD05). 
The GD05 models extend to 1 Myr ages through the use of theoretical 
libraries and allowed us to probe the full expected age range of the M33 
cluster system. The GD05 models also extend to older ages, but 
for ages greater than 63~Myr we prefer to use models based upon empirical rather 
than theoretical libraries.

Stellar population parameters were estimated using the ULySS (Koleva et al. 2009) 
IDL code which performs full spectrum fitting in pixel space to the observed spectrum 
using model templates.
The code minimises the $\chi^2$ between object and template represented by a linear
combination of non-linear model components.
We adopted a full spectral fitting approach instead of using index-index
plots in order to make full use of the spectral information at hand.
In comparison with indices, full spectral fitting can maximize 
the information available in the spectrum, at the potential expense of complicating 
the interpretation of the results and estimation of uncertainties.
This approach proved particularly useful for the younger clusters in our
sample. For ages $<$ 100 Myr by far the most prominent feature
in our optical spectra is the balmer H$\beta$ absorption line
(and HeI lines for very young clusters). Metal lines (iron, magnesium etc) 
are quite weak in integrated spectra at these young ages meaning
that individual measurements of these lines can have large uncertainties even 
for relatively high S/N spectra.

\begin{figure}
\centerline{\includegraphics[width=9.0cm]{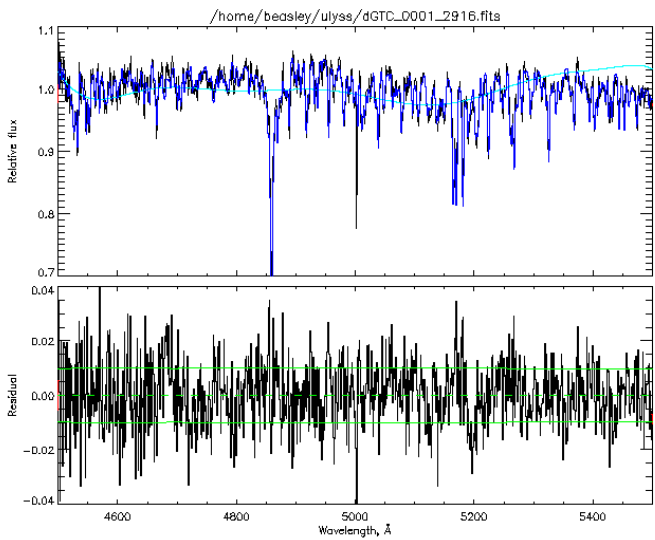} }
\caption{{\it Top panel:} comparison of the observed spectrum of cluster
S2916 (U49) (black line) compared to the best model SED (blue line).
The cyan curve represents the flux ratio between the models and observation.
{\it Bottom panel:} Residuals from division of the GTC spectrum
of the cluster by the best fitting SED. In this example of a good fit, 
residuals are of order $\sim2$ percent.  The green lines represent the 
mean and standard deviation of the residuals.
\label{ulyss_resid}}
\end{figure}

\begin{figure}
\centerline{\includegraphics[width=8.0cm]{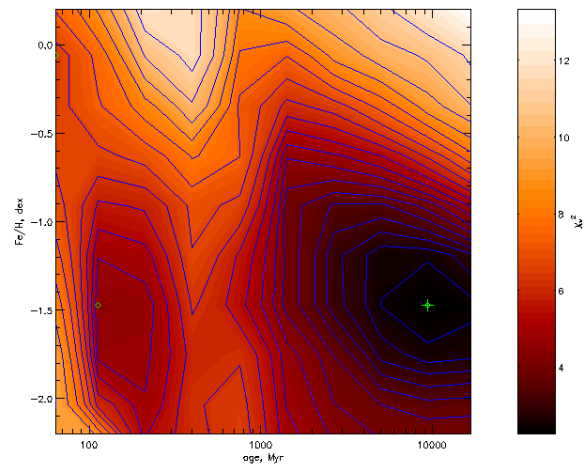} }
\caption{Map of $\chi^2$ values for the range of ages and metallicities
in the MILES models as determined by ULySS. The best solution lies at 
$\sim10$ Gyr ages and [M/H] = --1.3. However, another (``false minimum'')
solution lies at similar metallicities but $\sim100$ Myr ages. In this case ULySS
has correctly identified the true minimum. Metal-rich solutions at all ages
are rejected.
\label{ulyss_chimap}}
\end{figure}

For each cluster we allowed ULySS to search through a grid of MILES + GD05 SSP
models, calculate and locate the minimum $\chi^2$ as a function of age and metallicity.
Prior to these fits, we determined the ``line spread function'' of the GTC and WHT 
using several stellar templates observed as radial velocity standards (see Koleva et al. 2009).
We did not attempt to fit for $\alpha$-element abundances because the majority 
of star clusters are at young ages where spectral differences attributable
to varying [$\alpha$/Fe] are minimal (Vazdekis et al. 2015).
In addition, we did not explore in detail the possibility that some of the 
star clusters may have multiple stellar populations with different ages 
(e.g. Milone et al. 2015). We plan to explore this issue in future work. 
For each best-fit solution, we visually examined the spectrum, the residuals 
between the observed and model template spectrum, and  
the 2-dimensional $\chi^2$ maps produced by the code. 
In most cases, ULySS correctly identified the best-fit template and the 
residuals were of order of a few percent around most spectral features.
Examples of a $\chi^2$ map and the best solutions for the GTC spectrum of 
GC U49 are given in Figures~\ref{ulyss_resid} and \ref{ulyss_chimap}. 
ULySS correctly selects the true minimum based on the $\chi^2$ contours.

\begin{figure}
\centerline{\includegraphics[width=9.0cm]{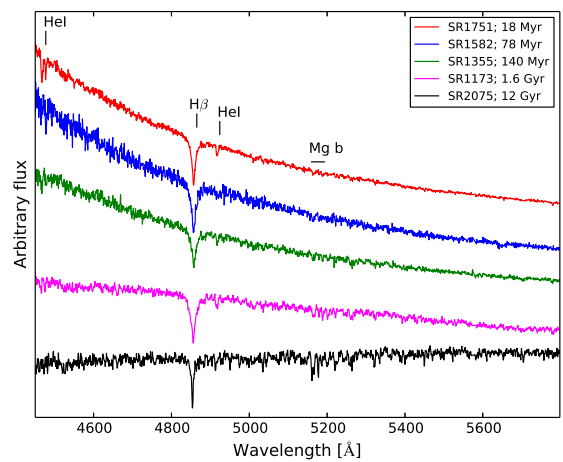} }
\caption{Example spectra of M33 clusters spanning a range of ages, 
from $\sim$20 Myr to 12 Gyr. The spectra have been offset by a constant
in flux for display purposes.Very blue continua, strong and broad
H$\beta$ lines and HeI lines are evident in the youngest clusters. 
Older clusters show redder continua and more prominent metal lines.
\label{example_spectra}}
\end{figure}

However, in some situations the H$\beta$ line was clearly over/under fit and checking
the $\chi^2$ maps showed that ULySS had settled either on a ``false minimum'', 
or had selected the true minimum but favoured a lower metallicity and 
a higher age than otherwise suggested by inspection of the H$\beta$ 
line (a manifestation of the age-metallicity degeneracy).
In the case of poor fits, we re-ran ULySS restricting the age-metallicity 
parameter space to what we considered best described the cluster under consideration
based on spectral features such as the width of H$\beta$, the presence (or lack of)
HeI lines and the continuum shape of the cluster.
Examples of our spectra with a range of ages are shown in Figure~\ref{example_spectra}.
In most cases, ages of $<$ 100 Myr were indicated by the presence of 
HeI lines ($\lambda\lambda$4712~\AA\ and 4920~\AA) originating in hot stars ($T\geq15,000$ K), 
and very blue continua with a few weak metal-lines
(we identified no clusters with HeII lines indicative of $\sim10$ Myr ages).
Older (100 Myr$-$1 Gyr) clusters also show broad H$\beta$ lines and blue continua,
but with increasingly strong metal lines (e.g., Mg at $\lambda\lambda5170~$\AA) and no HeI 
lines present. For clusters older than $\sim1$ Gyr, metal-lines strengthen further, 
the H$\beta$ line becomes increasingly narrow and the continuum reddens 
(to an extent dependent upon the metallicity of the cluster).
These considerations highlight the need for manual supervision while
applying this type of full spectrum fitting techniques to star cluster spectra.

Uncertainties in ages and metallicities were determined from Monte Carlo simulations
implemented in ULySS. For each spectrum, ULySS produces 100 synthetic 
spectra with S/N corresponding to the input spectrum generated using Poisson statistics 
and a random seed. 
ULySS was then run to find the best fit age and metallicity for each spectrum, 
and the uncertainties in age and metallicity were taken to be the standard deviation on the 
mean of the distributions of these parameters.
Typical uncertainties are $\sim30$ percent in age and $\sim0.2$ dex in metallicity for a median
S/N of $\sim30$ per~\AA, characteristic of this sort of analysis (see e.g., Wolf et al. 2007, 
As'ad et al. 2013). Note that the uncertainties given here account for the statistical 
errors based on the S/N of the spectra, and do not include systematic errors due to our 
choice of stellar population models.

\begin{figure}
\centerline{\includegraphics[width=10.0cm]{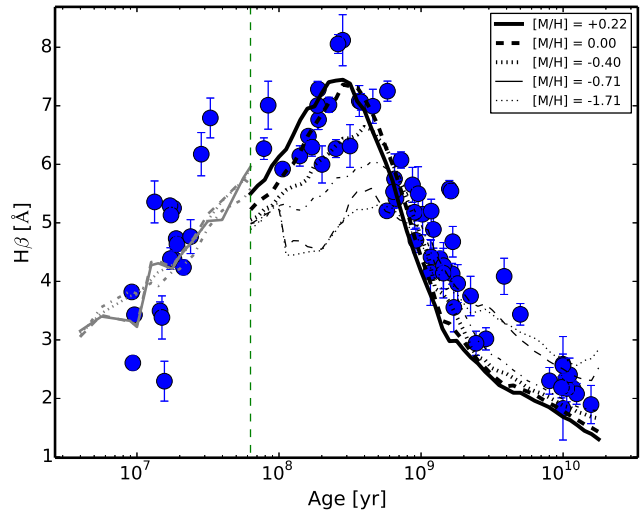} }
\caption{The LIS H$\beta$ index measured for the M33 clusters compared to 
our ULySS-derived ages. Models older than 63 Myr are those of MILES, 
those younger than 63 Myr (indicated by vertical dashed line) are those
of GD05. 
\label{hbeta}}
\end{figure}

One sanity check of our age determinations is to compare the ages 
from ULySS with those of the single-index model predictions for H$\beta$. 
This comparison is shown in Figure~\ref{hbeta}. Both the models
and spectra have been smoothed to a common dispersion of 5\AA\ in this 
case - the resolution for star cluster comparisons suggested 
in Vazdekis et al. (2010). 
The strength of the balmer lines in young stellar populations increase
as a function of increasing age until $\sim$400 Myr as A stars come to dominate the 
integrated spectrum (e.g. GD05). At older ages, the balmer lines then decline in strength 
as the main-sequence turn-off moves to lower temperatures. 
The models and data clearly show this behaviour Figure~\ref{hbeta}.
The cluster data follows the overall behaviour of the 
models as a function of age. Clearly the two measurements are not independent - the ages come 
from full spectrum fitting which include H$\beta$ which is often the 
strongest line in the integrated spectrum - but the agreement is
reassuring. In terms of the model predictions, there is a 
discontinuity at 63 Myr. This is the age representing the lower age 
limit of the Padova-based MILES models and where we use the GD05 models to predict ages and 
metallicities for the youngest clusters. 
A rescaling of the GD05 ages in this region
(for example) by $\sim15$ Myr to older ages would bring better 
agreement with the MILES models in terms of the H$\beta$ index, although we have not 
done this here since the origin of this offset is not clear\footnote{Both MILES and GD05 
use the Girardi et al. 2000 isochrones so the difference probably lies in the stellar 
libraries - empirical in the case of MILES, theoretical in the case of GD05.}.

Our final age and metallicity estimates with their associated 
uncertainties are given in Table~\ref{tab_sample}.

\section{Properties of the M33 cluster system}
\label{Results}

\begin{figure}
\centerline{\includegraphics[width=10.0cm]{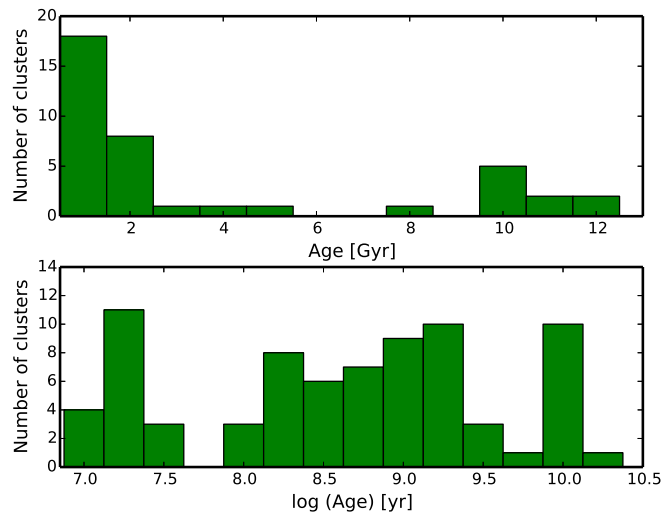} }
\caption{Distributions of the ages of the M33 cluster sample. More than
half of the cluster sample (47/77 clusters) are younger than 1 Gyr.
\label{agehist}}
\end{figure}

\begin{figure}
\centerline{\includegraphics[width=10.0cm]{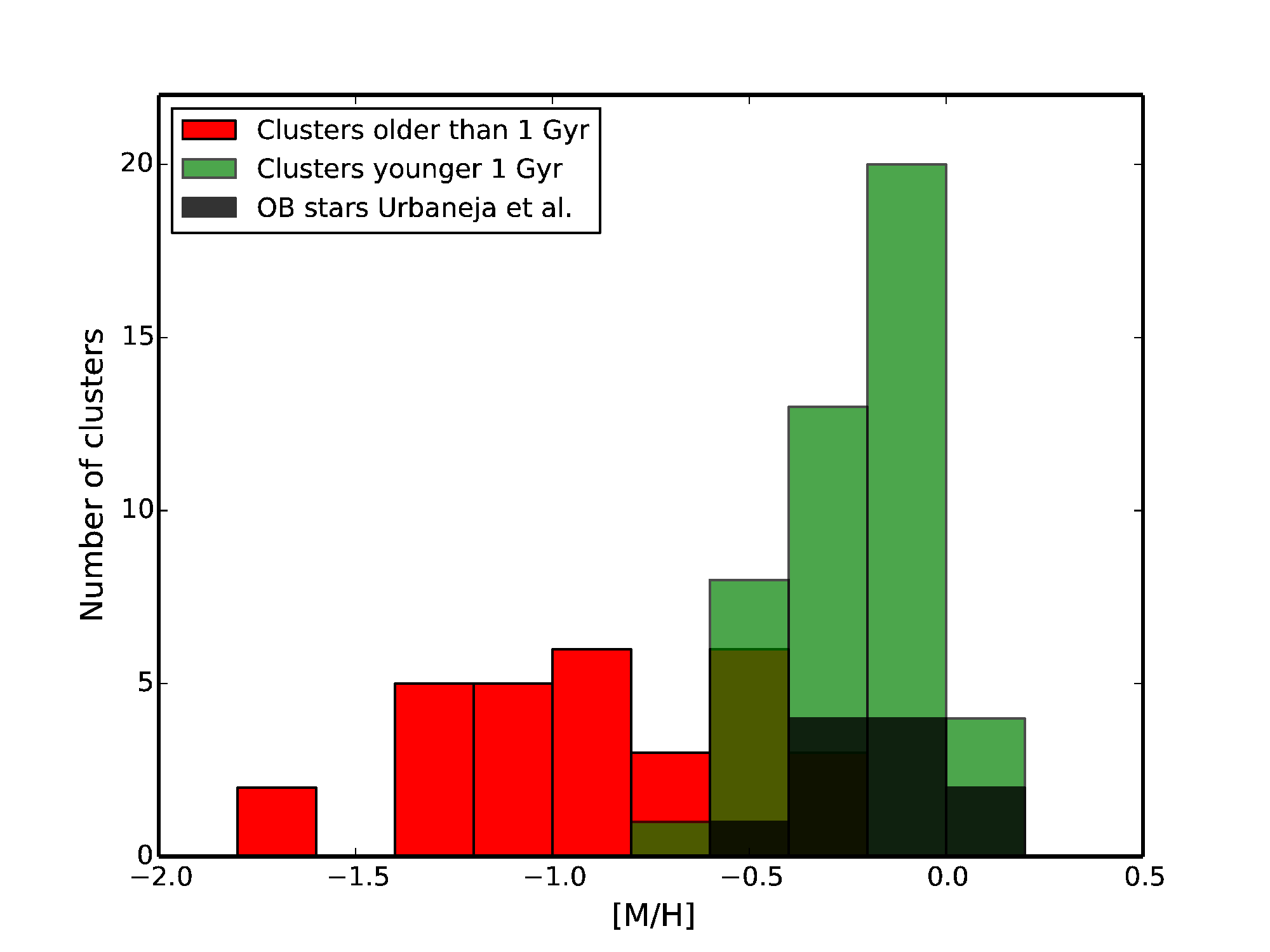} }
\caption{Metallicity distribution of the M33 cluster sample. 
Clusters younger than 1 Gyr are all more metal-rich than [M/H]=-1.0.
Also shown are the metallicities for M33 B-type supergiants from
Urbaneja et al. (2005). The massive stars and young clusters have similar
metallicities at or near the solar value.
\label{fehhist}}
\end{figure}

We first broadly characterised the M33 cluster sample based on their
stellar population properties. The distributions in age and metallicity
of the cluster samples, determined from our stellar population analysis, 
are shown in Figures~\ref{agehist} and \ref{fehhist}.
The clusters exhibit a range of ages, from very young $\sim10$ Myr
objects to $\sim13$ Gyr old globular cluster-like ages. Some 61 percent (47/77)
of the sample is younger than 1 Gyr. 
Based on the present data, M33 seems to have produced clusters throughout 
its lifetime. The relative lack of clusters between 3--9 Gyr may be 
due to an decrease in cluster formation at this epoch, or increase in cluster
destruction. A more complete spectroscopic sample will be required to explore the 
reality of this age gap.

In terms of metallicity (Figure~\ref{fehhist}), the clusters show a wide distribution
($-1.74<$[M/H]$<0.18$) with clusters younger than 1 Gyr being more  metal-rich ([M/H] $>-1.0$). 
This suggests a distinct age-metallicity relation for the clusters 
which we discuss in more detail in Section~\ref{subsec:AMR}. 
The metallicities of the youngest clusters are very 
similar to those found by Urbaneja et al. (2005)
for a sample of 11 B-type supergiants determined from spectral synthesis 
($\langle$[M/H]$\rangle = -0.14$, $\sigma_{\rm [M/H]}$ = 0.17). These massive stars with 
main-sequence lifetimes of $\sim$Myr constitute a metal-rich
disc population in M33. 
For very young clusters ($<100$ Myr), metallicities from integrated spectra 
are least well constrained since metallic features are weak 
compared to the Balmer series (this reduced metallicity sensitivity is 
indicated by the convergence of the lines of metallicity in the stellar 
population models at young ages-- see Figure~\ref{hbeta}). 
The good agreement in metallicity between the Urbaneja et al. (2005) sample
and the youngest M33 clusters is an important validation of our analysis
techniques.

\begin{figure}
\centerline{\includegraphics[width=10.0cm]{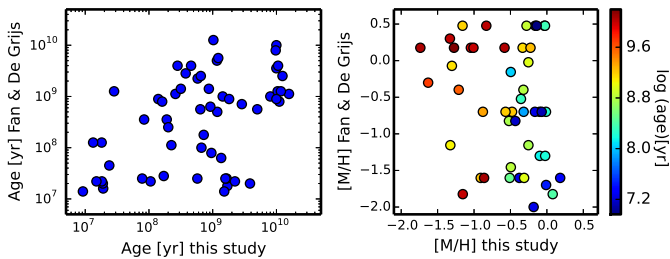} }
\caption{Comparison of our spectroscopic age ({\it left panel}) and 
metallicity ({\it right panel}) estimates for 58 clusters in common with 
the photometric estimates of  Fan \& de Grijs (2014).
The metallicities have been colour-coded based on our age determinations.
\label{compare_fan}}
\end{figure}

Fan \& de Grijs (2014) have presented ages and metallicities for a sample
of 671 M33 clusters based on fitting spectral energy distributions constructed 
from broad-band photometry. These authors identified a large number ($\sim140$)
of clusters with [Fe/H] $<-2.0$ projected onto the disc of M33.  
We see no clusters in our sample with [M/H]$<-1.74$. 
To investigate this further, we matched 58 clusters
in common between the present study and Fan \& de Grijs (2014) and compare  
the ages and metallicities of these clusters in Figure~\ref{compare_fan}.
The agreement between our ages and metallicities and those of 
Fan \& de Grijs (2014) is poor. In particular, there is no 
correlation at all between the cluster metallicities. There are 
clearly a number of very young, very metal-poor clusters in the  
Fan \& de Grijs (2014) sample not seen in our data. In addition, 
at any  given metallicity, the  Fan \& de Grijs (2014) show an
extremely wide range of ages ($10^{7} - 10^{10}$ yr) which is hard to 
understand both based on our results, but also in terms of the 
age-metallicity relations seen in Local Group dwarfs and the 
Magellanic Clouds (e.g., Carrera et al. 2011; Leaman et al. 2013).

Furthermore, there are six {\it bona fide} GCs in common in our
samples, five of which (M9, R12, R14, H38, U49 and U77) have been age-dated to be older than 
$\sim7$ Gyr via $HST$/WFPC-2 colour-magnitude diagrams (CMDs; Sarajedini et al. 2000).
We find ages older than 5 Gyr for all these GCs, while  Fan \& de Grijs (2014) 
obtain, with the exception of H38, younger ages
M9($1.2\pm0.2$ Gyr), R12($2.0\pm0.6$ Gyr), R14($4.0\pm0.7$ Gyr), H38($10.0\pm3.0$ Gyr), 
U49($1.2\pm0.2$ Gyr) and U77($0.6\pm0.1$ Gyr).
In view of the above, we conclude that the SED-fitting technique 
presented in Fan \& de Grijs (2014) does not break the age-metallicity
degeneracy in the sense that some very young clusters are identified 
as very metal-poor objects and some old, relatively metal-poor GCs 
are identified as young.

To return to the present sample, based upon the age distribution shown 
in Figure~\ref{agehist}, and for analysis purposes, 
we separated the clusters into subgroups based on their ages. We defined four 
subgroups: those that we termed {\it young} clusters ($<150$ Myr; $n$=22), 
{\it young-intermediate} age clusters (150 -- 1 Gyr; $n$=24), 
{\it intermediate} age clusters (1 -- 4 Gyr; $n$=19) and {\it old} (or globular) 
clusters ($>$4 Gyr; $n$=12). Age differences in these four subgroups might 
be expected to be reflected in their spatial, chemical and 
kinematical properties. 

\begin{figure*}
\centerline{\includegraphics[width=18.0cm]{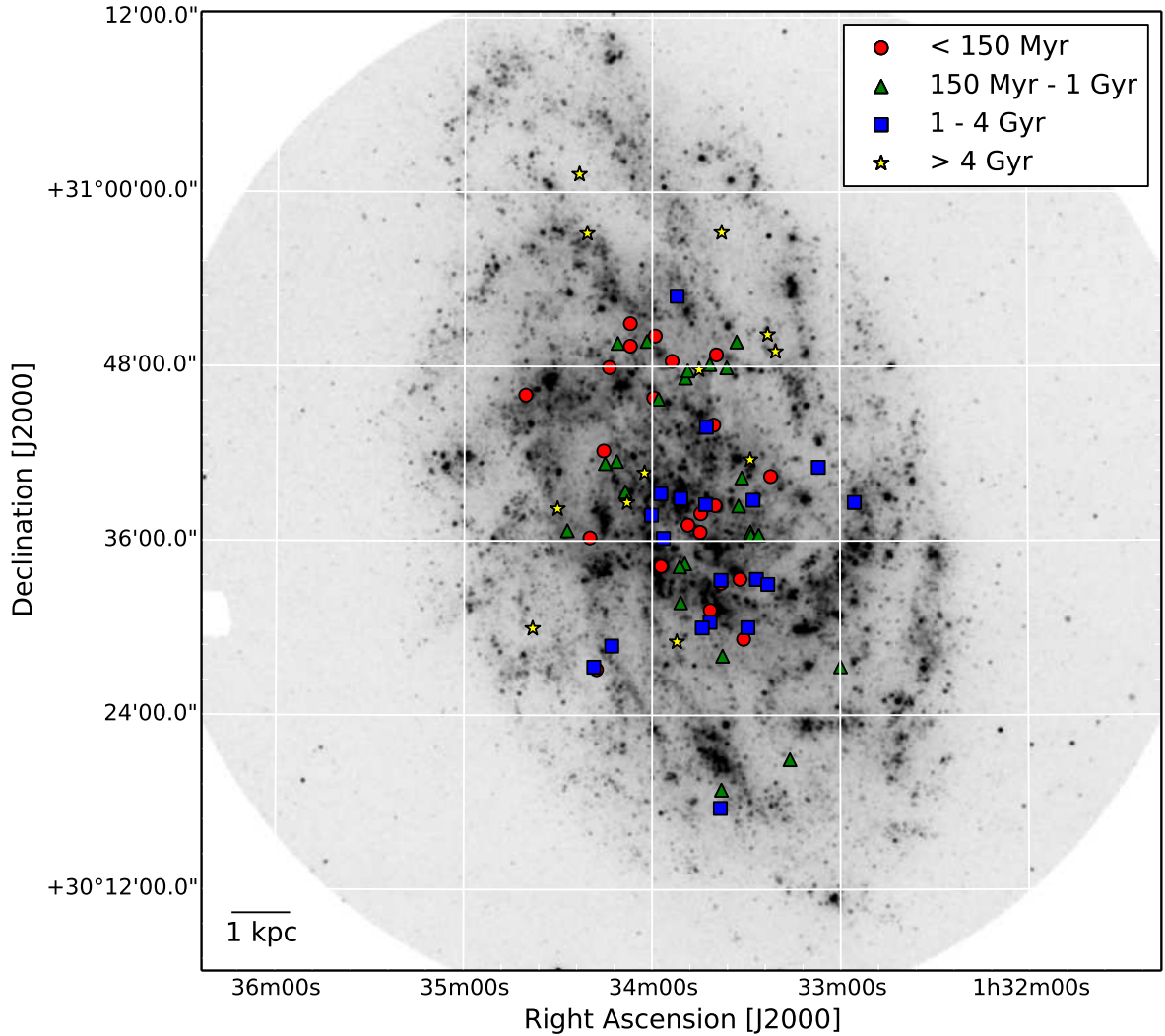} }
\caption{The projected spatial distributions of the sample of 
77 star clusters in M33 compared to a FUV GALEX image of the galaxy.
The clusters have been separated into four subgroups based upon
their ages {\it red circles :} $t<$150 Myr, {\it green triangles: } 150 Myr -- 1 Gyr, 
{\it blue squares :}  1--4 Gyr, {\it yellow stars:} $>$ 4 Gyr. 
Our sample is clearly confined (in projection) to the UV/optical disc of M33. 
\label{spatial_distribution}}
\end{figure*}

Figure~\ref{spatial_distribution} shows the 
spatial distributions of our four age subgroups compared with a 
far-UV (FUV; 1350-1750~\AA) GALEX image of M33.
Our spectroscopic sample is confined to within $\sim6$ kpc of the galaxy centre and 
therefore the following results relate to this inner region of the M33 disc.
Note, however, that a lack of clusters at larger radii is real in that 
all surveys of the M33 star cluster system to date have found that the cluster system
is more centrally concentrated than the field stars (e.g., Sarajedini \& Mancone 2007;
San Roman et al. 2010).

\subsection{Age-metallicity relations in the M33 disc}
\label{subsec:AMR}

\begin{figure}
\centerline{\includegraphics[width=10.0cm]{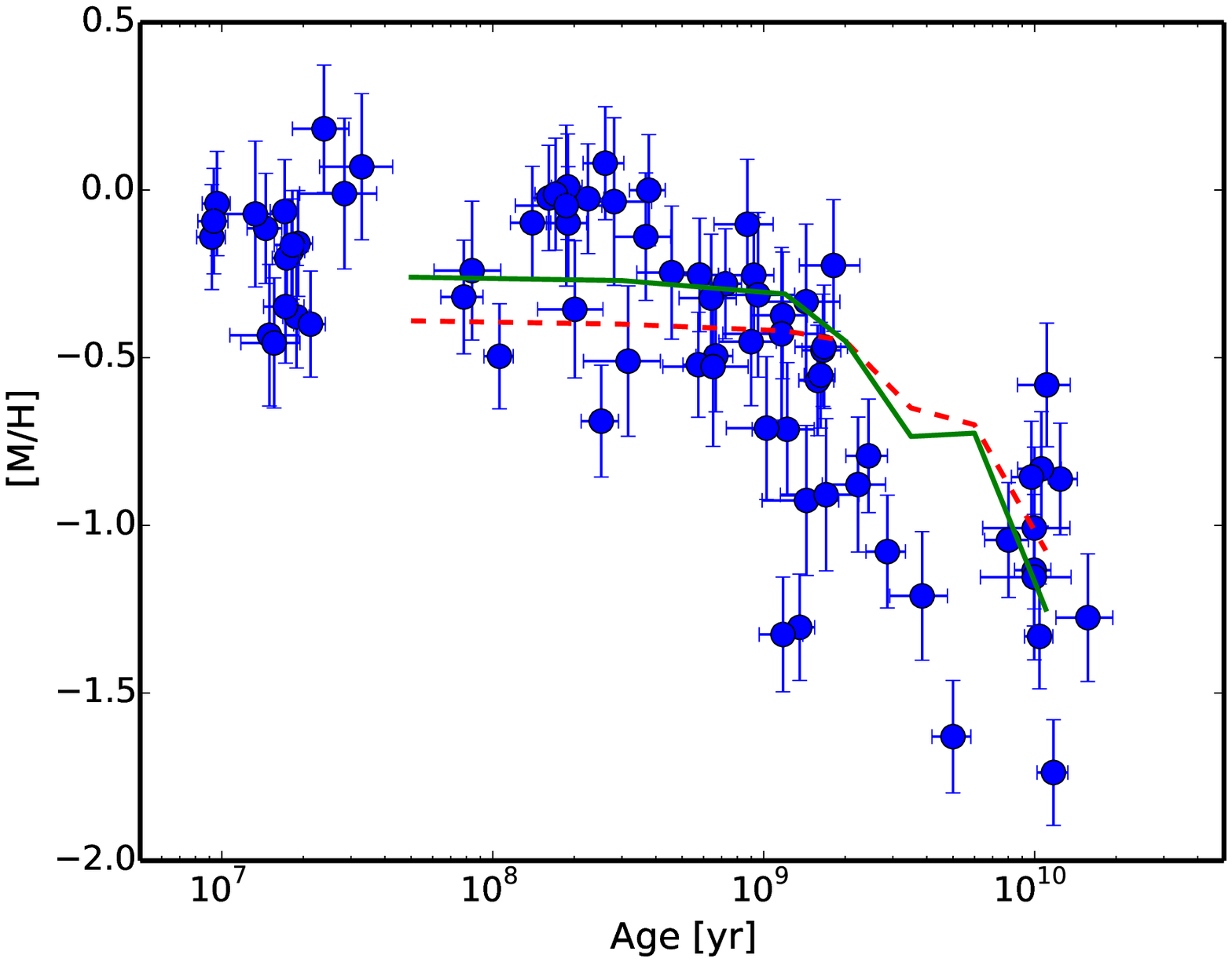} }
\caption{Ages and metallicities of the M33 clusters on a 
logarithmic age scale. Also shown are the ``free inflow'' chemical evolution
models of Barker \& Sarajedini (2007) for [M/H] and [Fe/H] (dashed and solid
lines respectively). 
\label{AMR_log}}
\end{figure}

\begin{figure}
\centerline{\includegraphics[width=10.0cm]{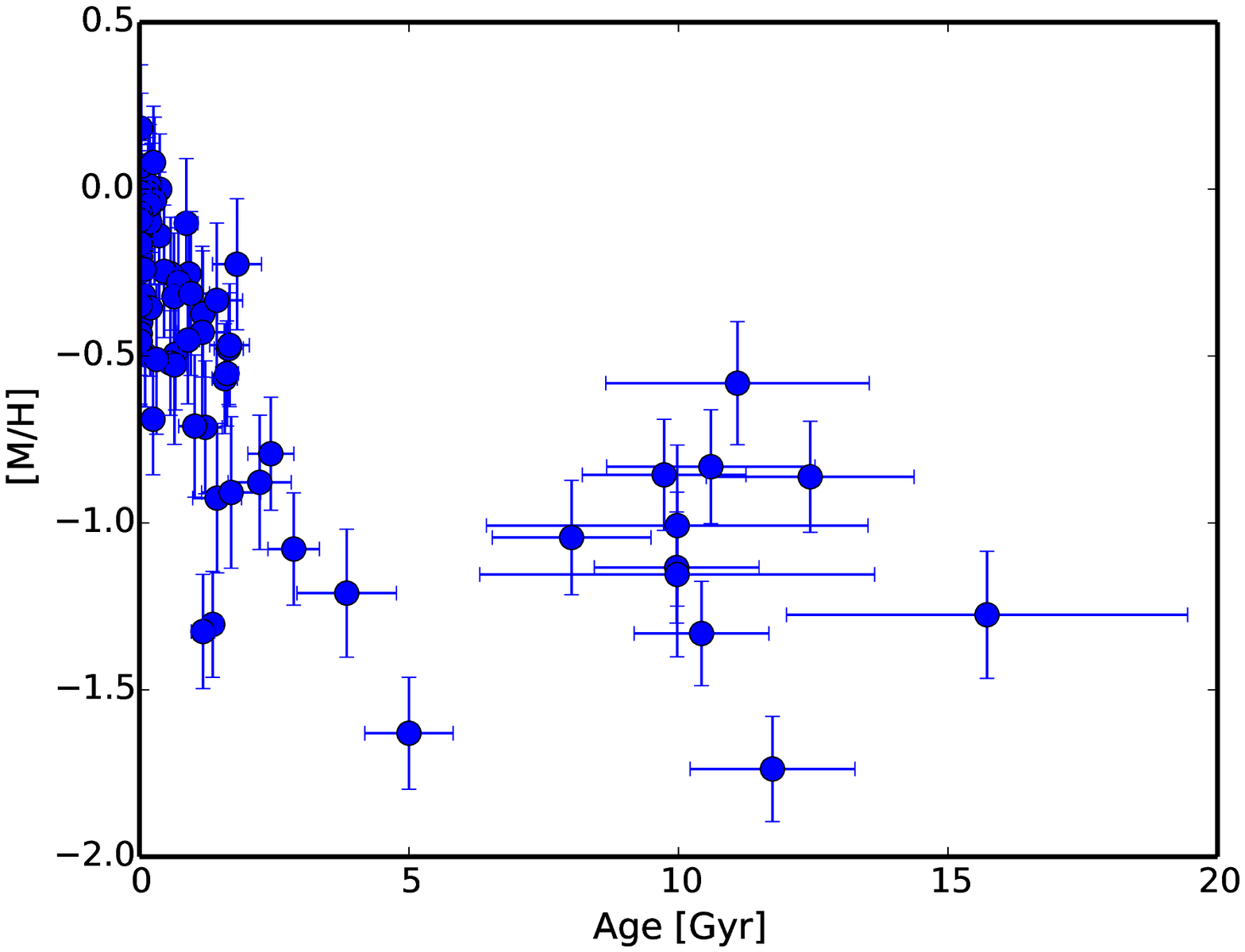} }
\caption{Ages and metallicities of the M33 clusters on a linear
age scale. 
\label{AMR_lin}}
\end{figure}

The AMR of stellar population is an important observational
constraint on chemical evolution models. Age-metallicity plots of 
the M33 clusters are shown in Figures~\ref{AMR_log} and~\ref{AMR_lin}. 
Mean ages and metallicities for the young, 
intermediate-young, intermediate and old subgroups are listed 
in Table~\ref{tab_grad}. The uncertainties in age and metallicity are 
the standard deviations on the mean values.

Figures~\ref{AMR_log} and~\ref{AMR_lin} show that there is a clear AMR in the cluster 
sample in the sense that older clusters are, on average, more metal-poor than younger clusters. 
Clusters younger than $\sim300$ Myr show an approximately
constant metallicity of $\sim$[M/H]=$-$0.2 
($\sigma_{\rm [M/H]}=0.17$)\footnote{It is not clear whether the apparent age gap at $\sim$50 
Myr is real, or rather reflects the difference in apparent age zero-points 
between the MILES and GD05 models.}. This is consistent with the
metallicities found for OB stars in the M33 disc (Urbaneja et al. 2005)
as mentioned previously.
Older clusters become progressively more metal-poor until 
the oldest clusters seem to either depart from the AMR,
or show a larger scatter in metallicity ($\sigma_{\rm [M/H]}=0.31$ for a given age)
than the younger clusters.
As we show in Section~\ref{Kinematics}, this seems to coincide with a 
transition from rotation-dominated kinematics to more dispersion-dominated 
kinematics.
Monte Carlo simulations indicate that the typical (statistical) 
uncertainty in our metallicities is $\sim$0.22 dex for both the MILES and GD05
stellar population models. We calculated the dispersions in metallicity of the
age subgroups (by first fitting and subtracting a 3rd order polynomial 
fit to the AMR) and find dispersions of 0.17 dex (young subgroup), 0.19 dex 
(intermediate-young), 0.31 dex (intermediate age) and 0.31 dex (oldest clusters).
Subtracting our observational errors in quadrature from these dispersions
we find that the intrinsic metallicity spread in the clusters
cannot be more than $\sim0.25$ dex at most, and for the two youngest 
cluster bins is equal to or less than our 0.2 dex metallicity uncertainties.

Also shown in Figure~\ref{AMR_log}
are chemical evolution models from Barker \& Sarajedini (2007)
which were explicitly modeled for the M33 disc. Qualitatively
the agreement between the model and cluster AMRs is reasonable,
although it is clear that for ages younger than 5 Gyr the modeled
AMR is flatter than the cluster AMR.
However, it should be noted that this comparison is not entirely appropriate since the 
Barker \& Sarajedini (2007) modeled an observed CMD $\sim$9 kpc
from the galaxy centre, whereas the majority of our clusters 
lie within 6 kpc from the centre of M33.

\subsection{Spatial distributions of the M33 clusters}
\label{subsec:spatial}

\begin{figure}
\centerline{\includegraphics[width=10.0cm]{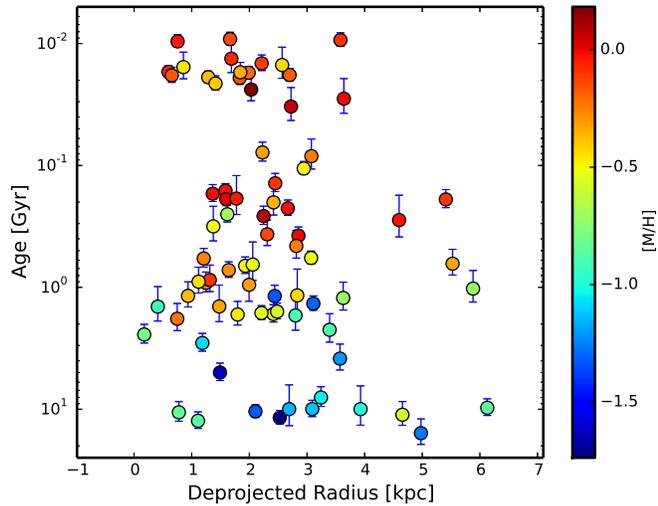} }
\caption{Ages of the cluster sample plotted as a function of galactocentric
radius. The clusters are colour-coded by their metallicities. 
\label{rad_age}}
\end{figure}

The deprojected radii of the clusters are shown as a function of their 
ages and metallicities in Figure~\ref{rad_age}. No clusters younger 
than $\sim100$ Myr are found beyond a projected radius of 4 kpc in our sample.  
More generally, the spatial distributions of youngest clusters show a mild 
tendency to be more centrally concentrated than the older clusters. 
The mean radius of each of the cluster age groups is given in Table~\ref{tab_rot} 
illustrating this trend.

\begin{figure}
\centerline{\includegraphics[width=6.0cm]{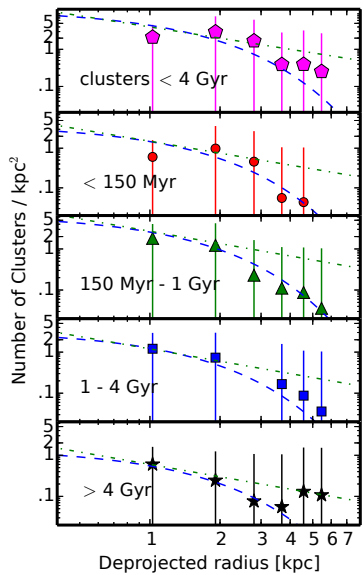} }
\caption{Surface density profiles of the four cluster subgroups using their
deprojected radii : {\it red circles :} $t<$150 Myr, {\it green triangles:} 150 Myr$-$1 Gyr, 
{\it blue squares :}  1$-$4 Gyr, {\it black stars :}  $>$4 Gyr. 
The surface density profiles for all clusters younger than 4 Gyr are shown as magenta pentagons.
Also shown are exponential ({\it long-dashed line}) and power-law
profiles with slope $-1$ ({\it dot-dashed line}) with arbitrary normalisation. 
\label{rad_profile}}
\end{figure}

Deprojected surface density profiles of the cluster subgroups 
are shown in Figure~\ref{rad_profile}. 
These density profiles are compared to an exponential profile ($N(R)\propto e^{-R}$), 
which might be expected for disc stellar populations, and a power-law relation 
($N(R) \propto R^{-1}$) the expectation for an isothermal sphere.
The clusters, in general, exhibit profiles which are more similar to exponential 
distributions than power-law distributions, i.e. the clusters have spatial distributions more
consistent with that expected of disc rather than spheriodal populations. 
Interestingly, and although limited by small numbers, the profiles for the two oldest 
cluster groups are suggestive of exponential profiles within $R\sim4.5$ kpc (i.e., within 
approximately two optical disc scale-lengths) but then appear to flatten beyond this radius.
The hint of an exponential surface density profile for the oldest 
clusters in our sample within $R\sim4.5$ kpc is both intriguing and unexpected. 
We return to this issue shortly.

\subsection{Metallicity gradients in the M33 disc}
\label{subsec:gradients}

\begin{figure*}
\centerline{\includegraphics[width=13.0cm]{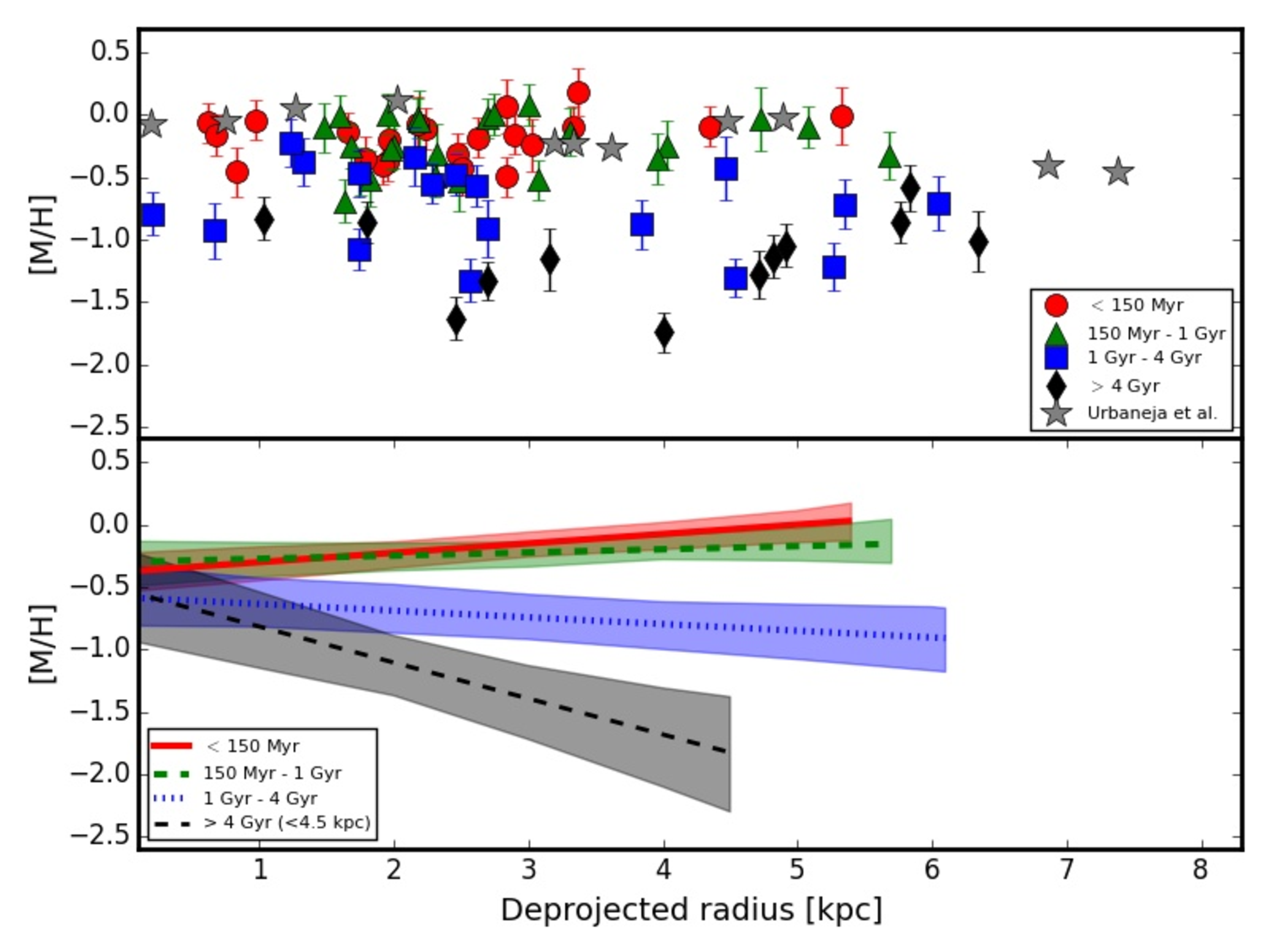} }
\caption{Radial metallicity gradients of the M33 clusters. 
{\it Top panel:} metallicities of the clusters as a function of deprojected 
galactocentric radius. The clusters are binned as a function of age:
{\it red circles:} $t<$150 Myr, {\it green triangles:} 150 Myr$-$1 Gyr, 
{\it blue squares:}  1$-$4 Gyr, {\it black diamonds:}  $>$4 Gyr.
{\it Lower panel:}  Gradients of the cluster subpopulations from linear
least-squares fits. The shaded regions represent the 10- and 90- percent
confidence limits obtained from bootstrap resampling. 
For the oldest clusters, only the slope and confidence limits for the inner ($R<$ 4.5 kpc)
clusters are shown (see text).
\label{rad_grad}}
\end{figure*}

The disc radial metallicity gradient, and any evolution 
(or lack thereof) of this gradient places important constraints on 
disc formation models (e.g., Chiosi 1980; Boissier \& Prantzos 1999; 
Cioni 2009; Marcon-Uchida, Matteucci \& Costa 2010; Gibson et al. 2013). 
We explored the M33 disc metallicity gradient as traced by its star clusters in our data.
Since the inclination and position angle of the semi-major axis of M33 is known, 
we can explore the behaviour of the cluster metallicities as a function of deprojected radius.
To deproject the clusters into the plane of the galaxy we assumed an inclination 
angle to the line of sight of $i=56$ degrees (Paturel et al. 2003) and a position angle for 
the major axis of $\phi=22$ degrees (Warner, Wright \& Baldwin 1973).
The metallicities of the three youngest cluster groups are plotted as a function
of deprojected radius in Figure~\ref{rad_grad}.

For each cluster subgroup we performed weighted linear least-squares fits
with radius as the explanatory variable. The results of these fits (gradient, intercept)
are given in Table~\ref{tab_grad}. We estimated confidence intervals 
(10 and 90 percentiles of the distributions) and uncertainties on the gradients of these 
fits by running 1,000 Monte Carlo (bootstrap with replacement) simulations. 
Our fits and confidence intervals are shown in the lower
panel of Figure~\ref{rad_grad}.
For the two youngest cluster subgroups we identify mildly positive radial 
metallicity gradients. The youngest subgroup has formally a steeper
positive gradient (d[M/H]/dR = $+0.045\pm0.033$ dex/kpc) than the 150 Myr$-$1 Gyr
group ($+0.024\pm0.035$). However, within our confidence intervals the 
two distributions are quite similar.
For clusters in the $1-4$ Gyr age group, we find a negative radial gradient 
($-0.054\pm0.046$ dex/kpc). I.e., the gradients of the cluster populations 
show a tendency to become less negative with time.
Combining the three youngest cluster groups, we find a gradient 
of ($-0.026\pm0.038$ dex/kpc) in the disc clusters. 
By comparison, Frinchaboy et al (2013) find a gradient of $-0.09\pm0.03$ dex/kpc 
for open clusters within a galactocentric radius of 10 kpc in the Milky Way.
Steeper gradients have been found by Urbaneja et al. (2005) for 
supergiant stars in M33 (our fits to the Urbaneja et al. data : $-0.057\pm0.025$). 
These gradients are also similar to 
the median [O/H] gradient measured for nearby isolated late-type spirals 
($-0.041\pm0.009$ dex/kpc; Rupke, Kewley \& Chien 2010) and are consistent with 
our data.

We also show metallicity as a function of radius for the oldest clusters ($>$4 Gyr)
in Figure~\ref{rad_grad}. A linear least-squares fit yields a gradient of 
d[M/H]/dR = $0.04\pm0.06$ dex/kpc for these clusters (for clarity 
we do not show the confidence limits on this fit in Figure~\ref{rad_grad}).
However, inspection of Figure~\ref{rad_grad} shows curious behaviour;
the inner clusters ($R<$4.5 kpc) exhibit a steep negative gradient, while
clusters beyond this show a flatter relation (with significant scatter).
The gradient for the $R<$4.5 kpc clusters is $-0.29\pm0.11$ dex/kpc.
That is, there appears to be a break in the metallicity gradient
of the oldest clusters at about $R\sim$4.5 kpc. 
Interestingly, this is also the location where the oldest clusters, 
and the  1$-$4 Gyr cluster group, show a hint of a transition from an 
exponential distribution to a flatter (power-law like) distribution 
(see Figure~\ref{rad_profile}). One possibility is that we are seeing 
a superposition of two old cluster populations in M33, a disc-like
inner population and a more spherically distributed outer population.
We return to this issue in more detail in Section~\ref{subsec:Globularclusterages}.

Returning to the M33 cluster sample as a whole, 
while larger samples would be desirable to confirm the trends, 
our data does suggest an evolution of the M33 metallicity gradient 
in the sense that the gradient has flattened (or rather, become less 
negative) with time.
This is in qualitative agreement with higher redshift systems. 
Not only are discs expected to undergo significant size evolution
with time (in the sense that discs are smaller at higher redshifts; 
e.g. Brooks et al 2010), also radial metallicity gradients of higher
redshift systems seem to be steeper than those observed locally.
For example Yuan et al. (2011) claim a measurement of 
a metallicity gradient for a lensed grand-design spiral at $z=1.49$
($\sim9.3$ Gyr for $H_0=70$, $\Omega_M=0.3$, $\Omega_\Lambda=0.7$)
based on H{\rm II} regions. These authors obtain an extremely steep
gradient in [O/H] ($-0.16\pm0.02$  dex/kpc) in a system with a
dynamical mass of $1.3\pm0.2 \times 10^{10}$~\msun (within 2.5 kpc). 
Interestingly, Jones et al. (2010) claim an even steeper metallicity gradient
in a lensed galaxy at $z=2$ (the so-called ``Clone arc'') finding
$-0.27\pm0.05$ dex/kpc in an even more massive system ($\sim 2.4~\times~10^{10}$~\msun, 
within 2.9 kpc)\footnote{Caveats include that these high-redshift systems will 
presumably evolve to be significantly more massive than M33. 
In addition, the high-redshift observations measure gas abundances whereas our 
gradients are derived from the metals already locked up in stars.}

\begin{figure}
\centerline{\includegraphics[width=7.0cm]{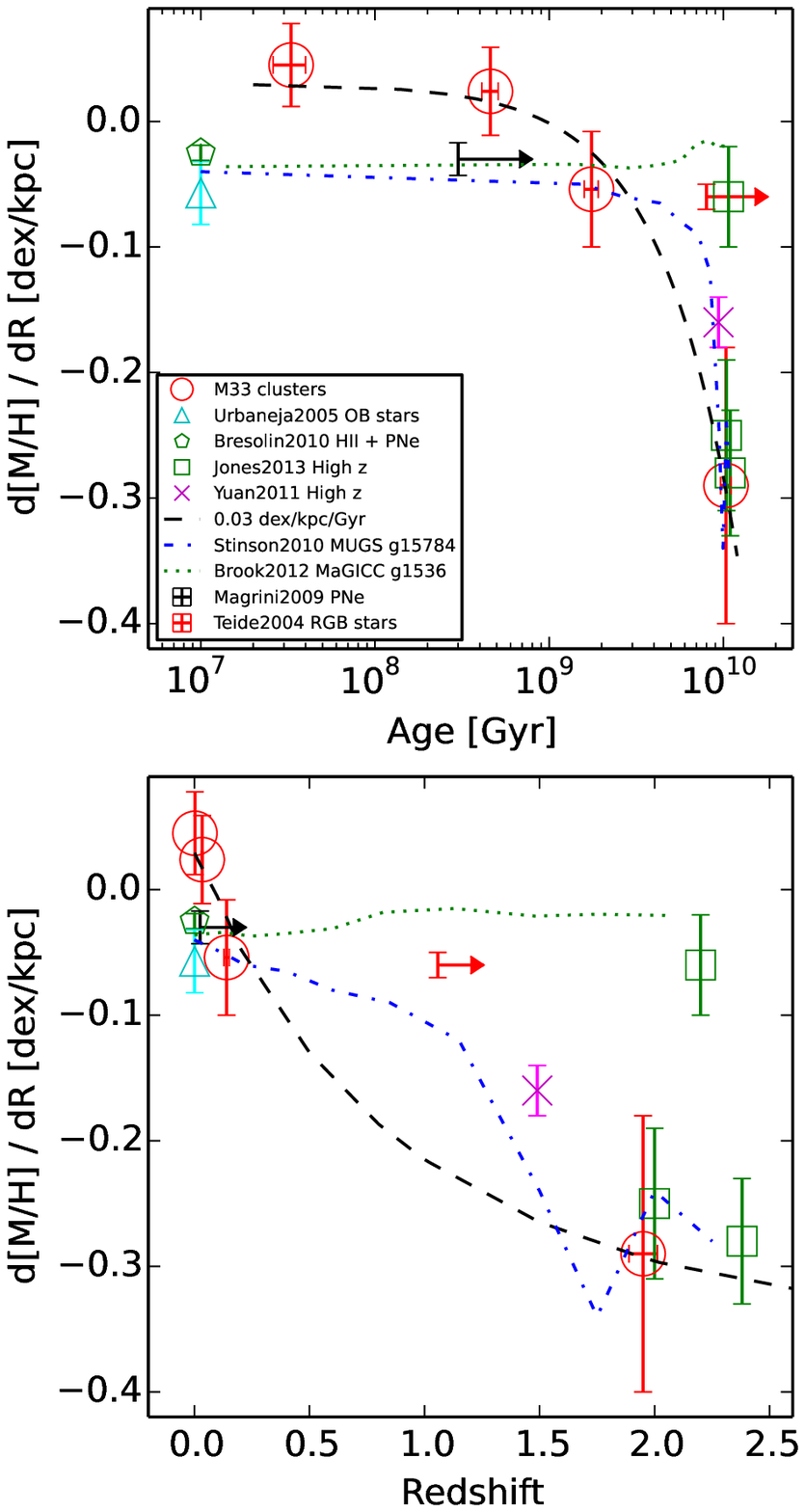} }
\caption{Time evolution of the metallicity gradient in M33.
Top panel shows gradients as a function of population age, 
lower panel as a function of redshift.
The M33 clusters (red circles) are compared to the gradients determined
from OB stars (Urbaneja et al. 2005), HII regions
and PNe (Bresolin et al. 2010; Margini et al. 2009), RGB star
photometry (Teide et al. 2004). Also shown are gradients determined for 
high-redshift systems (Jones et al. 2013; Yuan et al. 2010).
Predictions are also shown for two different simulations of
L$^{*}$ disc galaxies (Brook et al. 2012; Stinson et al. 2010).
The black dashed line shows a fit to the M33 data giving an 
evolution of the metallicity gradient of $0.03$ dex/kpc/Gyr.
In the top panel we have fixed the ages of the OB stars from Urbaneja et al. (2005)
and HII and ``young'' PNe from Bresolin et al. (2010) to 10$^{7}$ yr for
display purposes.
\label{gradz}}
\end{figure}

We compare the measured evolution of the M33 metallicity gradient for our
star cluster sample with data from other tracers of gradients in M33 and also
high redshift systems in Figure~\ref{gradz}. For M33, data for OB stars comes
from Urbaneja et al. (2005), HII and PNe from Bresolin et al. (2010) and 
Magrini, Stanghellini and Villaver (2009) and CMD-based estimates of metallicity from 
Tiede, Sarajedini and Barker (2004). Data for high-redshift discs comes
from Yuan et al. (2011) and Jones et al. (2013) (see above). Note that the emission-line
data (for HII, PNe and the high redshift systems) measure [O/H] rather than
[Fe/H] (or total metallicity [M/H]). We make the assumption that, in terms of gradients,  
these quantities track each other to first order. 

In terms of the M33 data there is some disagreement as to whether or not the 
gradient evolves with time. Our data suggests that it does - and quite strongly. 
The  dashed line in Figure~\ref{gradz} shows a fit to the M33 cluster data yielding
an evolution in the metallicity gradient, d[M/H]dt / dR = $0.03$ dex/kpc/Gyr.
This strong evolution is consistent with observations of some high-redshift
systems (Jones et al. 2013; Yuan et al. 2011). However, the RGB stars and PNe 
suggest a much weaker evolution, which is also consistent with some high redshift
observations (e.g., see the redshift $\sim2.2$ system in the data of Jones et al. 2013;
Figure~\ref{gradz}). This uncertainty is reflected in the simulations.
The results of two simulations for $L^{*}$ discs (i.e., a factor of $\sim10$ more 
massive than M33)  with differing feedback prescriptions are shown in
Figure~\ref{gradz}. The realisations come from the ``McMaster Unbiased Galaxy Simulations'' 
(MUGS; Stinson et al. 2010) and ``Making galaxies in a Cosmological Context'' 
(MaGICC; Brook et al. 2012; Stinson et al. 2013).

The MUGS simulations reflect the M33 data (and high-redshift systems) 
quite well, consistent with a strong evolution in the metallicity gradient. 
However, the MaGICC simulations - which show little evolution - better 
reflect the RGB and (possibly) the PNe data. 
The principal differences between the two simulations are that MUGS-g15784 uses a 
thermal feedback scheme that injects approximately 40 percent less energy into the 
ISM from supernovae than MaGICC (see Gibson et al. 2013 for more details). 

The principal observational problem in constraining the time-evolution
of metallicity gradients in local discs is age-dating older ($>2$ Gyr) stellar
populations. This is indicated by the lower limits on the 
RGB star and PNe data in Figure~\ref{gradz} and reflects the
poorly constrained ages of these evolved populations. This is less
of an issue for the youngest populations (e.g., HII, OB stars, young clusters) 
where the metallicity gradients agree quite well (c.f., Figure~\ref{gradz}).
Spectroscopic metallicities of RGB stars and more data on the 
intermediate-age to old clusters should help to better constrain this regime.

In summary, we find a flattening of the M33 metallicity gradient with time from
the star cluster data. This is in agreement with high redshift observations of
more massive discs, but in some tension with observations of RGB stars and PNe
in M33. Our results are consistent with the picture that the inner galaxy regions 
of the M33 disc is more chemically evolved than the outer regions. I.e., consistent 
with an ``inside-out'' disc formation scenario (e.g. MacArthur et al,. 2004; 
Jones et al. 2010). 

\begin{table*}
\caption{Stellar population properties measured for cluster subgroups.}
\begin{tabular}{lccccccccc}
\hline \hline
Subsample & mean age & mean [M/H] & $\sigma_{\rm [M/H]}$ & [M/H] gradient & intercept & N\\
 & (Gyr) & (dex) & (dex) & (dex/kpc) & (dex) & \\
\hline
young & $0.033\pm0.007$ & $-0.19\pm0.04$ & 0.17 & $0.045\pm0.033$ & $-0.377\pm0.019$ & 22 \\
young-intermediate & $0.46\pm0.05$ & $-0.23\pm0.04$ &  0.21 &  $0.024\pm0.035$ & $-0.300\pm0.020$ & 24\\
old-intermediate & $1.75\pm0.16$ & $-0.74\pm0.08$ & 0.33 & $-0.054\pm0.046$ & $-0.582\pm0.022$ & 19\\
old & 10.35$\pm$0.71 & $-1.12\pm0.09$ & 0.32 & $0.041\pm0.050$ ($-0.290\pm0.110$)$^{a}$  & $-1.282\pm0.010$ ($-0.530\pm0.048$)$^{a}$ & 12\\
\hline
$^{a}$ $R<4.5$ kpc.
\end{tabular}
\label{tab_grad}
\end{table*}

\subsection{Kinematics}
\label{Kinematics}

Previous studies (Schommer et al. 1992; CBFS02) have shown that the young M33 clusters rotate
in a similar sense to the HI disc, and that the average velocity dispersion of M33 cluster
populations increases with the age of the population. 
However, these studies were only able to place upper limits
on the velocity dispersions of the young clusters due to uncertainties on the individual
cluster velocities of 25-30 km/s, of order or larger than the HI disc dispersion
($\sim18.5$ km/s; Putman et al. 2009). In addition, age information in Schommer et al. (1992)
and CBFS02 came mainly from broadband photometry which is less able to disentagle
age-metallicity degeneracies in integrated stellar populations than integrated spectroscopy.

\begin{figure*}
\centerline{\includegraphics[width=13.0cm]{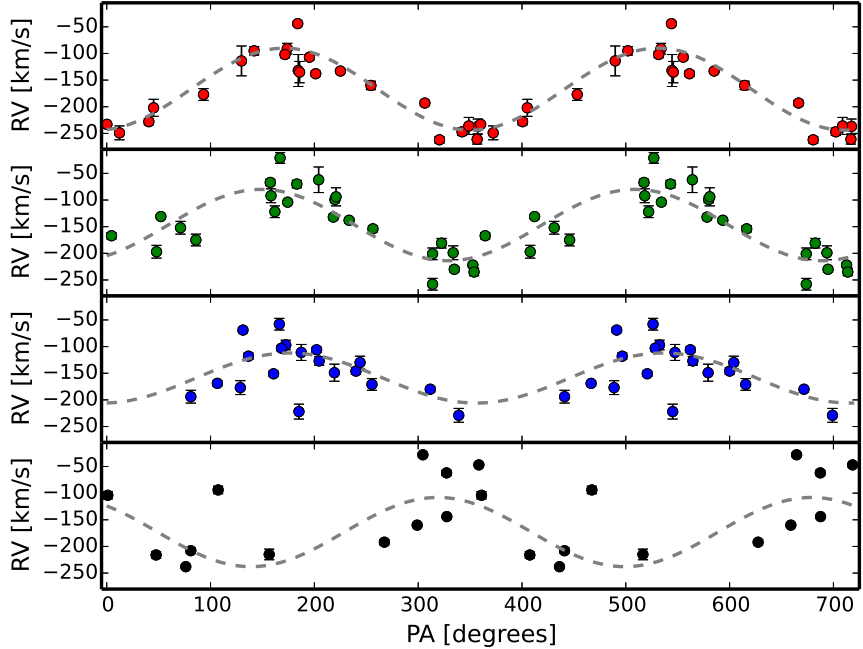} }
\caption{Radial velocities of cluster subsamples as a function of 
position angle with respect the major axis of M33 ({\it top to bottom:}
$t<150~Myr$, 150 Myr -- 1 Gyr, 1 -- 4 Gyr and $>4$ Gyr respectively).
The dashed ({\it grey}) lines are fits to these data as described in the text.
Note that the x-axis has been folded to better show the rotation in the system.
The velocity uncertainties are generally smaller than the symbol size.
\label{rot_plot}}
\end{figure*}

We show the radial velocities of our sample plotted as a function of position angle  
with respect to that of the M33 major axis in Figure~\ref{rot_plot}. 
Position angles  were calculated following the common convention E through N, adopting 
the NED value for the centre of M33 ($\rmn{RA}(2000)=01^{\rmn{h}}~33^{\rmn{m}}~50\farcs89$,
$\rmn{Dec.}~(2000)=30\degr~39\arcmin~36\farcs8$).
The clusters have been separated into the previously defined age bins. 
Rotation is clearly present in the three youngest cluster subgroups (Figure~\ref{rot_plot}), 
with the dispersion about the rotation solution increasing with the mean age of the 
subgroup. The oldest clusters also show some evidence for rotation, but with a rotation 
axis misaligned with that of the younger clusters. However, both the amplitude and position 
angle of this rotation is poorly constrained due to the small number of clusters in the 
oldest bin (see below).

To quantify the rotation amplitude and position angle in the cluster subgroups 
we assumed that the clusters are confined to a thin plane of inclination $i$ to the 
line of sight, and performed non-linear least-squares fits to the cluster bins using 
the functional form:

\begin{equation}
V({\rm \theta}) = V_{\rm sys} + V_{\rm rot}\times sin(\phi-\phi_{\rm 0}) sin(i)
\end{equation}

where $V({\rm \theta}$) is the observed cluster velocity at position angle $\theta$, 
$V_{\rm sys}$ is the systemic velocity of the clusters, $V_{\rm rot}$ the rotation
amplitude and $\phi$ the position angle of the line of nodes of the best-fit 
rotation solution with respect to that of the M33 major axis ($\phi_{\rm 0}$). 
We left $V_{\rm sys}$, $V_{\rm rot}$  and $\phi$ as free parameters to be 
determined by the fits. Uncertainties were determine by Monte Carlo 
simulations (1,000 bootstrap resamples with replacement for each age bin). 
We also tested our rotation solutions by fixing the position angle of the line of nodes 
to that of the youngest cluster bin (81$\pm9$ degrees) which is coincident with 
the M33 minor axis. The rotation amplitudes do not change significantly 
when fixing the position angle.

Our fits are shown graphically in Figure~\ref{rot_plot} and listed in 
Table~\ref{tab_rot}. We calculated the line of sight velocity dispersion of the cluster
subgroups by subtracting the best-fit rotation curve and then measuring 
the velocity dispersion of each group. These values are also given in Table~\ref{tab_rot}.
In addition, we include a small correction for our intrinsic velocity uncertainties which  
are subtracted in quadrature from the velocity dispersion estimates 
(a correction of $\sim2$ km/s). It is evident that there is a trend for the rotation 
amplitude to decrease and velocity dispersion
to increase with increasing cluster age (as quantified by $V/\sigma$ in 
Table~\ref{tab_rot}). A similar effect was identified by CBFS02.

We interpret the decrease in the rotation amplitude as a function
of group age as a manifestation of asymmetric drift; the increasing tendency 
of older populations to ``hide'' their rotation in randomized motions. 
Therefore, we corrected the rotation velocities of our cluster groups
for asymmetric drift by assuming 
a flat rotation curve and that the star clusters follow an 
exponential surface density profile (Figure~\ref{rad_profile})
(see Hinz et al. 2001) leading to the expression (from Leaman et al. 2012):

\begin{equation}
V^{2}_{\rm cor} = V^{2}_{\rm rot} + \sigma^{2}_{\rm los} (2 \frac{R}{R_{d}}-1)
\end{equation}

where $V_{\rm cor}$ and $V_{\rm rot}$ are the asymmetric-drift corrected
and observed, inclination-corrected velocities of the cluster subgroups
respectively,  $\sigma_{\rm los}$ the velocity dispersion of the clusters, 
$R$ the mean radius of the cluster group and $R_{d}$ the disc scalelength.

These corrected rotation velocities are listed in Table~\ref{tab_rot}.
Once corrected for asymmetric drift, the rotation amplitudes for all four cluster subgroups 
are similar ($\sim100$ km/s).
This is also true for the oldest cluster subgroup (although note that the uncertainties 
here are of order $\sim70$ percent of the rotation amplitude). 
The rotation of all the clusters is consistent with the rotation curve of M33 (Corbelli \& Salucci 1999) 
at the mean radii of the cluster subgroups.

Another approach to quantify the $V/\sigma$ of the clusters 
is to compare the cluster kinematics to the HI rotation curve.
CBFS02 did this by creating a disc model from the rotation
curve of Warner et al. (1973). Here, we take advantage of the available 
two dimensional data by measuring the HI velocity 
directly at the location of each cluster from the VLA HI 21cm velocity 
moment maps of Gratier et al. (2010). These data are available
at spatial resolutions of 25, 17 and 12 arcseconds respectively.
We chose to use the 25 arcsecond maps due to their more complete 
spatial coverage over the M33 disc, although the use of different resolution maps
does not significantly affect our results.

\begin{figure}
\centerline{\includegraphics[width=10.0cm]{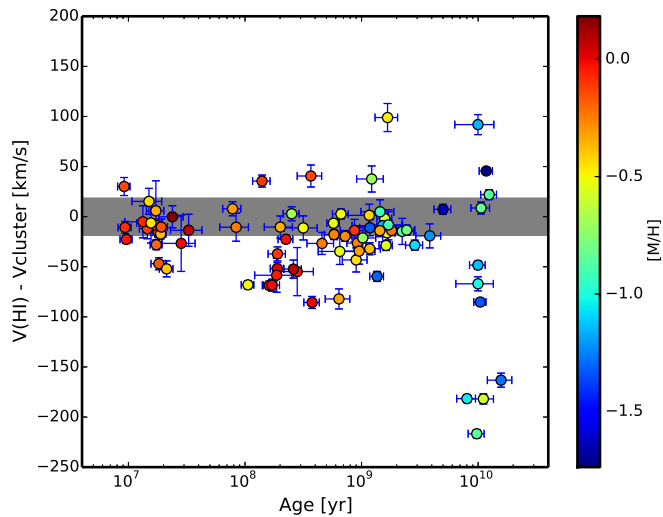} }
\caption{ $V_{\rm HI} - V_{\rm cluster}$ versus age for the M33 clusters. 
The horizontal shaded region represents the mean velocity dispersion ($\sim18.5$ km/s) 
of the HI in the star forming disc (Putman et al. 2009).
\label{vdisk}}
\end{figure}

The quantity $V_{\rm HI} - V_{\rm cluster}$, a measure of the deviation of the 
cluster velocity from the mean HI gas velocity at the position of the cluster, 
is plotted against cluster age in Figure~\ref{vdisk}. 
The velocity dispersions about the HI gas disc
for the four (youngest to oldest) subgroups are : 23.7$\pm5.0$, 29.1$\pm6.0$, 31.1$\pm8.0$ and 
98.2$\pm25.0$ km/s respectively. These values are consistent with our previous results 
from our simple disc model. The three youngest cluster groupings clearly 
constitute disc populations and the clusters show an increasing velocity dispersion with 
age. The youngest population has dispersion 23.7 km/s which is close to  the mean 
HI disc dispersion (18.5 km/s; Putman et al. 2009). Creating an additional bin with clusters 
younger than 50 Myr (N=18), yields a mean value for  $V_{\rm HI} - V_{\rm cluster}$
of 19$\pm$5 km/s.

\begin{table*}
\caption{Kinematical solutions obtained for cluster subgroups. V$_{\rm rot}$ indicates the rotation amplitude of 
the cluster subgroups corrected for inclination. V$_{\rm cor}$ indicates V$_{\rm rot}$ corrected for asymmetric drift. }
\begin{tabular}{lccccccccc}
\hline \hline
Subsample & V $\sin i$ & V$_{\rm rot}$ & V$_{\rm cor}$ & Position angle & Vsys & $\sigma_{\rm los}$ & V$_{\rm rot}$ / $\sigma_{\rm los}$ & Mean radius\\
 & (km/s)  & (km/s) & (km/s) & (degrees) & (km/s) & (km/s) & (km/s) & (kpc) \\
\hline
young & $77\pm11$ & $95\pm14$ & $100.4\pm29.1$ & $81\pm9$ & $-167\pm8$ & $20.0\pm5.0$ & $4.8\pm1.4$ &  2.38 \\
young-intermediate & $67\pm9$ & $83\pm11$ &  $103.6\pm25.0$ & $59\pm9$ & $-147\pm7$ & $29.5\pm6.0$ & $2.8\pm0.7$ &  2.74 \\
old-intermediate & $47\pm18$ & $58\pm22$ &  $85.9\pm38.0$ & $86\pm20$ & $-159\pm15$ & $38.9\pm8.9$ & $1.5\pm0.7$ &  2.85 \\
old & $65\pm40$ & $80\pm49$ & $158.5\pm107.2$ & $229\pm44$ &  $-173\pm35$ & $67.3\pm19.4$ & $1.2\pm0.8$ & 3.96  \\
\hline
\end{tabular}
\label{tab_rot}
\end{table*}

\subsection{The age-velocity dispersion relation}
\label{AVR}

\begin{figure*}
\centerline{\includegraphics[width=12.0cm]{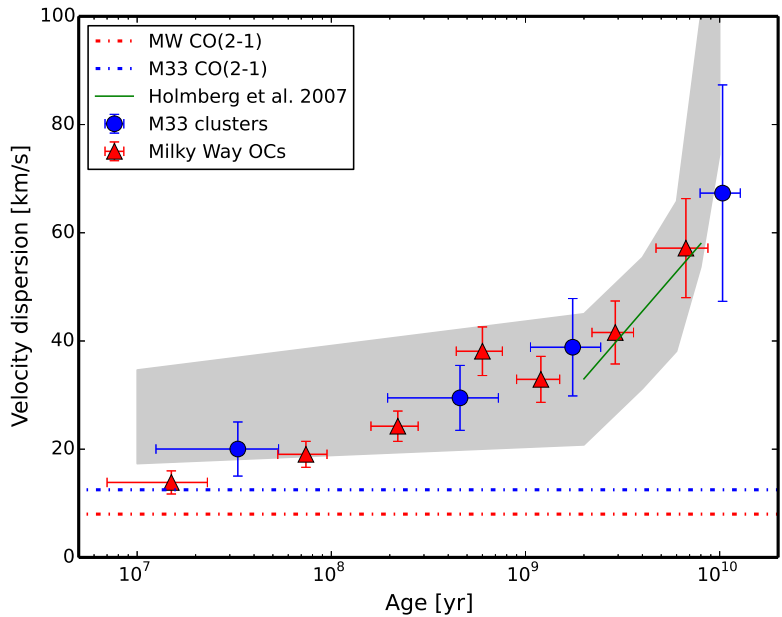} }
\caption{Age-velocity dispersion relation of M33 clusters. 
Velocity dispersion of M33 clusters versus age compared to 
Milky Way open clusters (134 clusters; Hayes \& Friel 2014). Also shown are estimates of the 
inter-cloud molecular gas velocity dispersion for the Milky Way from
Stark (1984) and for M33 from Druard et al. (2014). 
The shaded region indicates the predictions from the simulations
of Martig et al. (2014). The AVRs for the Milky Way and M33 clusters 
look very similar.
\label{disp_age}}
\end{figure*}

The age - velocity dispersion relation (AVR) for the M33 clusters (with dispersions
taken adopting the thin-disc model) is shown in Figure~\ref{disp_age}. 
In the figure we also compare the M33 cluster data with 
the total LOS velocity dispersions of solar neighbourhood stars (Holmberg et al. 2007)
and those of Milky Way open clusters (OCs) compiled by Hayes \& Friel (2014).
In the case of the Hayes \& Friel (2014) data, we have multiplied their velocity
dispersions by a scalefactor of $\sqrt{3}$ assuming isotropic velocity distributions.
This is an estimate of the true scalefactor, since the Hayes \& Friel (2014) 
velocities contain both components of $\sigma_{U}$ and $\sigma_{V}$. 
However, the good agreement between the Holmberg et al. (2007) and Hayes \& Friel (2014) 
LOS dispersions for disc populations at a given age (after applying our scalefactor) 
suggests that this scaling is reasonable.

It is evident that the M33 clusters in Figure~\ref{disp_age} show an increasing 
velocity dispersion with age. This behaviour is similar to that 
of the Milky Way OCs. This is also observed in Milky Way disc stars, where 
the youngest stars in the solar neighbourhood have velocity dispersions 
of $\sim$10 km/s (again assuming isotropic velocity ellipsoids; Aumer \& Binney 2009), rising
to $\sim60$ km/s at 10 Gyr ages (e.g., Holmberg et al. 2007).

A number of workers have interpreted the AVR of the stellar disc 
with time as a result of disc heating processes (e.g. Spitzer \& Schwarzchild 1951; 
Wielen 1977). 
Candidate mechanisms for this heating may be secular - such as stellar interactions 
with giant molecular clouds (Spitzer \& Schwarzchild 1951 - but see Lacey (1984))
and spiral arms (Minchev \& Quillen 2006), or related to substructure infall including
black holes from the Galactic halo passing through the disc (H\"anninen \& Flynn 2002) 
and the infall of satellites in the form of minor mergers within the context of the hierarchical clustering 
paradigm (e.g. Mihos \& Hernquist 1996; Benson et al. 2004; Villalobos \& Helmi 2008; 
House et al. 2011).

An alternative (and not mutually exclusive) interpretation of the AVR is that stars and 
star clusters of a given age may have been formed in gas with intrinsically different 
kinematical properties (e.g. Brook et al. 2004; Bournard, Elmegreen \& Martig 2009; 
Bird et al. 2013). 
Specifically, the average (molecular) gas velocity dispersion in discs
may have been higher at early times (e.g., F\"orster Schreiber et al. 2009;
Wisnioski et al. 2014), thereby leading to stellar populations formed at these epochs
with characteristically higher velocity dispersions. In some sense, 
this may be regarded as a ``cooling'' of the disc ISM over time.

Both heating and cooling might be expected to occur in discs, 
but teasing out the relative importance of these processes 
is difficult with these data. Indeed, even with full phase-space information 
and detailed chemical abundances this is a challenging task (e.g., Bovy et al. 2012).
However, the fact that we see similar AVRs in both the Milky Way and M33
discs does place constraints on the underlying cause of this relation 
(see also Leaman et al. 2013). 

Simulations that take into account both secular heating and merging processes make a 
fair job of reproducing the observations. In Figure~\ref{disp_age} we show the 
region of age-velocity dispersion space produced by the simulations of 
Martig, Minchev \& Flynn (2014)\footnote{Again we apply a $\sqrt{3}$ scalefactor, 
since Martig et al. (2014) provide $\sigma_z$ not $\sigma_{\rm los}$.}.
These authors re-simulated at high resolution 7 disc galaxies taken from 
cosmological simulations (Martig et al. 2012) with a range of formation
histories (quiescent to 1:4 mergers) and stellar masses ($3.3-19.1\times~10^{10}$\msun).
Although the stellar mass of the M33 disc is ($\sim3\times10^{9}$\msun; Corbelli 2003), 
which is a factor of 10 less massive than Martig et al's least massive galaxy, both the 
normalisation and trend of the simulations are broadly consistent
with the M33 and Milky Way cluster data at all ages (the simulations show a wide range 
in velocity dispersion since we have taken the minimum and maximum values of all 7 galaxies 
as the range in model predictions). In detail, the slope of the AVR in the simulations
is somewhat flatter than the observations. In addition,  the youngest Milky Way OCs 
have velocity dispersions that lie below the models. The latter discrepancy 
may reflect the resolution limits in the simulations resulting in stars being 
born too hot (Martig et al. 2014).

Although Martig et al. (2014) could not identify the exact mechanism(s) 
responsible for disc heating in their simulations, they conclude that for quiescent systems 
disc heating produces a smoothly increasing velocity dispersion up to 
$\sim$9 Gyr. A general result is that stars older than this are born kinematically 
hot in turbulent gas at early times (Brook et al. 2004; Bournard, Elmegreen \& Martig 2009; 
Bird et al. 2013). 

The M33 data seems consistent with the picture of a smoothly heated 
disc cluster population, perhaps with the oldest population formed in a kinematically 
hotter environment. However, the similarity between the AVRs of the Milky Way OCs and 
the M33 clusters  suggests that something other than just secular heating is controlling the 
overall shape of the AVR. Analytic models (Leaman et al. 2015) that account for both substructure 
heating and ISM evolution as a function of host galaxy mass suggest that the ISM defines 
a characteristic ``pressure floor'', with additional heating via subtructure playing 
an increasingly important role for systems with masses similar to, or greater than, that of 
M33. 

Finally, we highlight that comparisons between the AVRs of stars and star clusters in 
a given disc should place constraints on heating processes within stellar discs. 
We expect that the efficiency of mechanisms that heat a population via scattering will 
be a function of the mass ratio of the scatterer and ``scatteree'' 
(e.g. Spitzer \& Schwarzchild 1951). For example, in scenarios where stars and star clusters 
are heated via scattering off GMCs in the plane
of the disc, we would expect stars of a given age and position in the disc to be more efficiently 
scattered than star clusters, yielding a higher line of sight velocity dispersion.

\subsection{The globular clusters}
\label{Theglobularclusters}

We identify 12 clusters in our sample that have age, metallicity and kinematical
properties consistent with globular clusters in M33.
Six of these GCs in our sample (R12, R14, M9, U49, H38, U77) have $HST$ CMDs 
(in F555W and F814W filters) that extend to $\sim1.5$ magnitudes below the horizontal branch 
(Sarajedini et al. 2000). The remaining six clusters we identify as good GC candidates, 
having radial velocities and stellar population properties consistent with this interpretation 
(Section~\ref{StarClusterIdentification}).

\subsubsection{Globular cluster ages}
\label{subsec:Globularclusterages}

A number of previous studies have concluded that there may be a significant age spread 
amongst the M33 GCs based on integrated broadband photometry 
(Cohen, Persson \& Searle 1984; CBFS02), integrated spectra 
(Christian \& Schommer 1983; Chandar et al. 2006) or arguments based 
on the horizontal branch morphology (Sarajedini et al. 2000). 
Sarajedini et al. (2000) found that the clusters R12, U49 and H38 have 
exclusively red horizontal branches (red clumps), despite their relatively low metallicities.
Based on this fact, Sarajedini et al. concluded that these clusters exhibit the 
so-called second-parameter effect, and argued that under the assumption that the 
second parameter is age then R12, U49 and H38 may be as young as $\sim$ 7 Gyr. 
We find old ages for these three clusters ($\sim10$ Gyr). 
Given the uncertainties in the modelling of integrated stellar populations and the 
continued controversy over what drives the second parameter effect (e.g. Milone et al. 2014) 
we do not consider that our ages are in serious disagreement with the 
Sarajedini et al. (2000) study.

Sarajedini et al. also identified the clusters M9 and U77 as being genuinely old
($\sim12$ Gyr) clusters. For M9 we obtain an age of $\sim12$ Gyr in good agreement with  
Sarajedini. However, for U77 we obtain an age of $\sim6$ Gyr
even though our derived metallicities are identical within the uncertainties 
with Sarajedini et al. (see below). 
The $HST$ CMD of U77 clearly shows stars blueward of the HB instability strip 
suggesting that this is indeed an old cluster even though we find younger age solutions 
from ULySS. Previous studies indicate that blue HB stars can affect integrated spectra
to the extent that old stellar populations can look young if the HB is not 
properly accounted for in SSP models (e.g. Beasley et al. 2002; Schiavon et al. 2004; 
Yoon, Yi \& Lee 2006; Cenarro et al. 2008). The MILES models model a canonical blue HB 
at low metallicities, but an excess of hot blue stars may be artificially lowering the 
spectroscopic age of this cluster. 
We indentify two further GC candidates with 
ages similar to U77. CMDs will be very valuable for investigating whether these are 
genuinely younger objects, or exhibit blue HBs. For example, Chandar et al. (2006) 
conclude, based on MMT integrated spectra and $HST$ photometry of the horizontal branch, 
that the M33 cluster C38 (not in our sample) has ages $\sim$2--5 Gyr. 
The observed lack of a blue HB in C38 supports this claim.

CBFS02 also give ages for five clusters in common with our 
sample (R12, R14, M9, U49-2916 and H38)  which were based on integrated colours (CBFS99). 
CBFS02 determined ages : R12 (2.5 - 10 Gyr), R14 (10 - 25 Gyr), M9 (1.25 - 2 Gyr), 
U49 (1.25 - 2 Gyr) and H38 (1.25 - 2.5 Gyr) (where the age ranges come from the error 
bars given by CBFS02). From our analysis we find that all of these
clusters are old with an age range between 9.5 -- 12.4 Gyr, this is consistent with the 
CBFS02 findings for R12 and R14, but inconsistent with M9, U49 and H38.
Due to the difficulties of breaking the age-metallicity degeneracy with broadband colours, 
we prefer our age estimates over those of CBFS02.

To summarise our age results, we find no compelling evidence, based on the present data, 
for a significant age spread in the GCs of M33 as claimed by previous authors. 
However, our sample is small, and ages from integrated
spectra lack fidelity at old ages and can also be affected by non-canonical 
hot stellar populations. Only deep CMDs reaching to below the main sequence 
turnoff in these clusters will provide secure ages for these objects.

\subsubsection{Globular cluster metallicities}
\label{subsec:Globularclustermetallicities}

\begin{figure}
\centerline{\includegraphics[width=10.0cm]{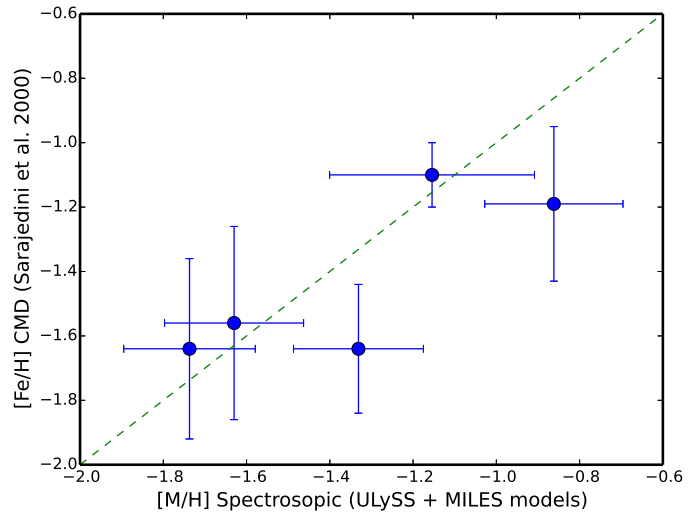} }
\caption{Comparison between our spectroscopic metallicities and those
determined by Sarajedini et al. (2000) based on the slope of the 
red giant branch from HST CMDs. The dashed line is a one to one relation.
\label{comp_sara}}
\end{figure}

Our spectroscopic metallicities for five clusters compared to
those derived by Sarajedini et al. (2000), based on the slope of the red 
giant branch, are shown in Figure~\ref{comp_sara} (R14 is not included due to high 
differential reddening in this cluster). We find reasonable agreement between the 
spectroscopic and CMD-derived metallicities in the range $-1.8 <$[M/H] $<-0.8$.

\begin{figure}
\centerline{\includegraphics[width=10.0cm]{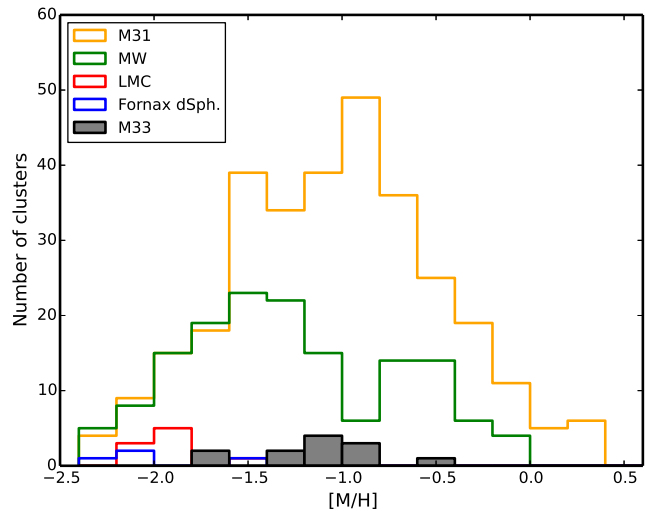} }
\caption{Distributions of the metallicities for old globular clusters 
in M31 (Caldwell et al. 2011), the Milky Way (see Harris 1996), 
M33 (this work), LMC clusters (Olszewski et al. 1991; Beasley et al. 2002)
and the Fornax dwarf spheroidal (Strader et al. 2003; Larsen et al. 2012).
\label{comp_met}}
\end{figure}

The metallicity distributions of the M33  GCs are compared to those 
derived for the Milky Way, M31, LMC and Fornax dwarf GCs in Figure~\ref{comp_met}.
These metallicities come from a variety of techniques; the majority of 
the Milky Way clusters have high dispersion spectroscopic measurements 
of individual stars (see Harris et al. 2006). Metallicities for M33 and 
M31 come from integrated spectra (this paper and Caldwell et al. 2011 respectively), 
the LMC clusters have metallicities predominantly from Calcium Triplet
spectroscopy of individual stars (Olszewski et al. 1991) and 
the Fornax dwarf spheroidal GC metallicities come from high resolution 
(Larsen, Brodie \& Strader 2012) and low resolution (Strader et al. 2003) integrated 
spectroscopy.
The metallicities from these varied approaches are consistent at
the $0.2-0.3$ dex level (e.g., Beasley et al. 2002; Puzia et al. 2002; 
Gonz\'alez Delgado \& Cid Fernandes 2010; Colucci et al. 2013).

Figure~\ref{comp_met} shows that the metallicities of the M33 GCs are quite
metal-rich ($\langle$[M/H]$\rangle$=$-$1.12$\pm$0.09)\footnote{If we consider 
only the M33 GCs identified by HST CMDs, the mean GC metallicity still remains high,
$\langle$[M/H]$\rangle$=$-$1.25$\pm$0.14}. This metallicity 
is entirely consistent with the expected location of M33 ($M_V\sim-18.9$) on the ``zero-age''
luminosity--metallicity relation derived by  C\^ot\'e, Marzke \& West (1998).
The mean M33 GC metallicities are also similar to the the mean metallicity of the 
M31 clusters ($\langle$[M/H]$\rangle$=$-$1.00$\pm$0.02), 
but more metal-rich than the Milky Way ($\langle$[M/H]$\rangle$=$-$1.27$\pm$0.05), 
and significantly more metal-rich than the LMC ($\langle$[M/H]$\rangle$=$-$1.87$\pm$0.08) 
and Fornax dwarf spheroidal ($\langle$[M/H]$\rangle$=$-$2.08$\pm$0.17) cluster systems.
Restricting the Milky Way clusters to the metal-poor halo population  
($\langle$[M/H]$\rangle$=$-$1.53$\pm$0.04) makes the apparent disparity with M33 more acute.
A similar separation cannot easily be performed for the M31 cluster system since the 
metallicity distribution of its cluster system is not clearly bimodal, although 
Caldwell et al. (2011) do identify a possible metal-poor peak at [M/H]$\sim-1.4$.

We investigated further the differences between the Milky Way and M33 clusters by 
determining, via Monte Carlo simulation, the likelihood that the M33 GCs could be 
drawn from a parent population resembling the Milky Way halo clusters. 
We cut both samples at [M/H] <=-0.7 (i.e. restricting 
ourselves to the halo clusters) giving 11 M33 clusters and 118 Milky Way 
clusters. We then ran 10,000 trials where we randomly selected 11 clusters 
from the Milky Way distribution and calculated their mean metallicities. 
Only 0.24 percent of these samples had mean metallicities equal to or greater than 
the mean of the M33 clusters. Therefore, based solely on metallicity, the M33 clusters 
are unlikely to have been drawn from a parent population resembling the Milky Way halo clusters
at $\sim99.8$ per cent confidence.

\subsubsection{On the origin of the globular cluster population}
\label{subsec:Origin}

The high metallicity of the M33 GCs presents something of a problem if
we wish to identify the majority of these objects solely with an accreted halo 
population (e.g. Searle \& Zinn 1978; C\^ot\'e, Marzke \& West 1998). 
The mean metallicities of LMC and Fornax clusters suggest that no combination 
of the GC systems of these (present day) galaxies could have contributed to the 
build up the metallicity distribution of M33 GCs in our present sample, save 
perhaps for the very metal-poor tail of the distribution.
Since Fornax (M$_{*}\sim10^{8}$\msun; Mateo 1998) and the LMC (M$_{*}\sim10^{9}$~\msun) bracket 
the lower and higher mass ends of Local Group ``dwarfs'' with GC systems, this 
constrains the nature of the origin of the M33 GCs. 

Assuming a Fornax dwarf or Small Magellanic Cloud (SMC)-like field star AMR, 
we can ask what age a GC accreted from such a 
galaxy need have in order to attain a metallicity of [M/H]$\sim-1.1$, the mean metallicity 
of the M33 GC system. The AMR for the SMC field stars suggests that 
an age of $\sim5$ Gyr or younger would be required to achieve the necessary enrichment levels
(Dobbie et al. 2004; Carrera et al. 2008; Leaman et al. 2013).
Such young ages for the GCs in our sample are generally ruled out based on our stellar 
population analysis and previous observations (see below). 
This age constraint can be relaxed to $\sim10$ Gyr 
if we assume that a more massive, LMC-like system were accreted. 
However, given that the LMC stellar disc is similar
in mass to the M33 stellar disc ($\sim3\times10^{9}$~\msun), such an equal-mass merger
would likely disrupt the entire M33 disc. A system intermediate in mass between the SMC
and LMC could possibly accommodate both the metallicities and ages seen in the M33 GCs, 
but some fine-tuning would be required (since the bulk on the inner M33 GC system would have
to be built up of such systems). 

Interestingly, the simulations of Cooper et al. (2010) 
suggest that the typical satellite that went to form a {\it Milky Way}-like halo had a mass 
similar to that of the brightest Local Group dwarf spheroidals (dSphs) 
(e.g. Fornax-like systems). However, differences observed between the stellar populations 
of halo stars and those seen in Local Group dSphs (Venn et al. 2004; Helmi et al. 2006; 
Fiorentino et al. 2014) suggest that these systems played at most a minor role in building 
up the Milky Way halo. If the M33 clusters truly represent the remnants of an 
accreted population then these clusters must have come from systems more massive than presently 
observable dSphs. 

An alternative origin for some or all of these GCs is a formation during an early 
collapse phase, perhaps leading to the formation of a bulge or disc component (e.g. Brook et al. 2004).
The question of whether or not M33 has a bulge remains controversial 
(e.g. Hodge 2012). However, Regan \& Vogel (1994) placed an upper limit on any putative 
bulge of M$_V>-16$ (L$_V<2\times10^{8}$~\lsun). Assuming a relatively generous $V$-band
specific frequency of 1 (Harris et al. 1991), this suggests that at most 3 GCs may be 
associated with this component. An alternative is to associate the GCs with the 
M33 disc. This association is not unreasonable; our GC sample is confined to within $\sim6$ kpc
of the galaxy centre and are seen in projection against the optical disc 
(Figure~\ref{spatial_distribution}). The inner ($R<4.5$ kpc) GCs show a 
hint of an exponential-like surface density profile and also a steep and negative
radial gradient consistent with the evolution of the gradient seen for 
the younger disc clusters. As discussed above, the M33 clusters in our sample 
are more metal-rich than Milky Way halo clusters, and also appear to be more metal-rich than 
expected for true halo clusters in M33 (e.g. Stonkut\`e et al. (2008) find [M/H]$\leq-1.4$ for an 
extended star cluster lying 12.5 kpc in projection from the centre of M33).
In addition, the kinematics of the clusters are consistent with a continuous
heating of the disc cluster population (or a cooling of the ISM from which they 
formed - see Section~\ref{AVR}) over the disc lifetime (Section~\ref{Kinematics}).

Finally, clues to the origin of the inner M33 GCs comes from analysis of their
structural parameters. Mackey \& van den Bergh (2005) have shown that the  
``young halo'' (YH) GCs in the Milky Way tend to have larger half-light
and tidal radii than the inner ``old halo'' (OH) clusters, which in turn are more
extended than the more metal-rich bulge/disc (BD) globular clusters.
Ma (2015) has presented structural parameters for 10 M33 globular
clusters (from Sarajedini et al. 1998), five of which are in our 
spectroscopic sample. All these clusters (R12, R14, M9, U49 and H39)
appear quite compact for their luminosities, with mean half-light
radii of $\sim3.5$ pc (King model) and a maximum $R_{h}\sim5.8$ pc (U49). 
This is similar to the BD clusters in  Mackey \& van den Bergh (2005) which 
all have $R_{h}<7$ pc. The distribution of half-light radii for the OH and 
YH clusters extends to $\sim17$ and 25 pc respectively.

Similarly, the mean tidal radii of the M33 
clusters is $\sim31$ pc with R14 having the largest ($R_{t}\sim40$ pc).
The BD clusters in Mackey \& van den Bergh (2005)
all have $R_{t}<60$ pc, with mean radii of $\sim30$ pc, consistent
with the M33 sample. Such small tidal radii are expected of clusters that spend a significant 
fraction of their time in a deep potential (i.e. near the galaxy centre), 
suggesting that these are truly central clusters and not simply halo
clusters seen in projection. By comparison, the OH and YH GCs in the Milky Way
have distributions of tidal radii that extend out to $\sim200$ and 150 pc respectively.

To summarise, while our sample size is small, based on their stellar populations, 
spatial locations, kinematics and structural parameters, we suggest that the 
inner M33 GCs are better associated with the M33 disc rather than its 
halo\footnote{Our sample is restricted to within $\sim6$ kpc of the galaxy centre. 
The very distant clusters identified in wide-field surveys (Huxor et al. 2009, 
Cockcroft et al. 2011) are probably genuine halo clusters. 
In addition, there may well be a mix of halo and disc clusters in our sample.} or minimal bulge.

\section{Conclusions}
\label{Conclusions}

We have obtained precision velocities (to $\sim10$ km/s) and stellar population parameters
(age, metallicity) for a sample of 77 star clusters in M33. Our principal findings are:

\begin{itemize}
 
\item The M33 disc clusters show a clear age-metallicity relation in the sense that 
younger clusters are more metal-rich than the older clusters. The youngest clusters 
in our sample have $\sim$10 Myr ages and near solar metallicities, very similar
to the M33 OB star populations.

\item We find evidence for evolution in the disc metallicity gradient in M33.
The metallicity gradient becomes less negative with time. The evolution
in the the metallicity gradient is strong with $\sim0.03$ dex/kpc/Gyr.
The inner globular clusters ($R<4.5$ kpc) show a steep, negative radial 
metallicity gradient.

\item We find little evidence for radial age gradients in the disc clusters. 
Clusters have continued to form throughout the star forming disc (within 6 kpc) 
throughout the disc lifetime.

\item Clusters younger than $\sim4$ Gyr all exhibit rotation consistent 
with the disc of M33.  The rotation amplitude decreases and velocity dispersion 
increases with increasing cluster age. We find a smoothly increasing
age-velocity dispersion relation very similar to that seen in the Milky Way
open cluster system. We interpret this as a combination of secular heating processes
and cooling of the ISM with time.

\item We identify six new globular cluster candidates that have kinematics and stellar
populations consistent with genuine globular clusters. Follow-up high resolution 
imaging is required to unambiguously determine their nature.

\item We find no strong evidence for a significant age spread in the M33 globular 
clusters. The majority of globular clusters in our sample are old ($\sim10$ Gyr).
Three clusters have spectroscopic ages of $\sim6$~Gyr. However, one of these is clearly older 
than this based on the presence of a well-developed blue HB in its $HST$ CMD 
(Sarajedini et al. 2000). The remaining two clusters may be genuinely intermediate 
aged clusters in M33.

\item The mean metallicity of the M33 globulars is relatively metal-rich 
($\langle$[M/H]$\rangle$=$-$1.12$\pm$0.09). This is significantly more metal-rich
than the halo GCs in the Milky Way, and GCs in the LMC and the Fornax dSph. Based on their
high metallicities, spatial distributions, kinematics and structural parameters we argue that the inner GCs 
are better associated with the M33 disc than halo.

\end{itemize}

We believe that this contribution reinforces the utility of star clusters
in studying the disc components of galaxies. Since clusters typically
span a large range of ages, they can be used to probe disc properties over the whole
disc lifetime, while at the same time consistent analysis methods can be applied.
This contrasts with complementary approaches that use stellar population tracers 
that are confined to narrow age ranges (e.g. OB stars; Urbaneja et al. 2005) or age ranges
that are not well constrained (e.g., PNe; Magrini et al. 2004; Ciardullo et al 2004) 
and require instrinsically different analysis methods.

The stellar population analysis used here, consisting of full spectral fitting 
(e.g., Koleva et al. 2009) to the latest generation of model SEDs (Vazdekis et al. 2010), 
is a powerful technique for estimating ages and metallicities of star clusters. 
However, it does have its limitations. Most apparent
is its lack of age resolution at old ages. Given the uncertainties in stellar population
modelling, particularly in understanding non-canonical hot populations and the 
incorporation of non-solar abundance ratios, our ability to differentiate age
amongst the oldest populations is limited. This is most apparent in our discussion
of the putative age-spread amongst the M33 globular clusters 
(Section~\ref{Theglobularclusters}). Only deep CMDs reaching to below the turnoff in 
these clusters will provide secure ages for these objects. 

Future improvements to this work should include increasing the sample size and spectral
range covered for the cluster populations. In particular, going to bluer wavelengths will
give additional leverage in constraining the ages of the younger clusters.
In addition, providing that the systematics are adequately characterised (Sakari et al. 2014), 
new techniques that obtain heavy- and light-element abundances via integrated spectroscopy 
(e.g. McWilliam \& Bernstein 2008; Colucci et al. 2012, 2013; Larsen et al. 2012; 
Sakari et al 2013) could usefully be applied to M33 star clusters and other nearby disc 
cluster systems. 

\section{Acknowledgements}

MAB thanks Tomas Ruiz Lara, Mina Koleva, Ryan Leaman, John Beckman, Chris Flynn, Bradley Gibson
and Marie Martig for useful discussions. 
This article is based on observations made with the Gran Telescopio 
Canarias (GTC), installed in the Spanish Observatorio del Roque de los Muchachos 
of the Instituto de Astrof\'isica de Canarias, on the island of La Palma.
This research has made extensive use of the NASA/IPAC Extragalactic Database (NED) which 
is operated by the Jet Propulsion Laboratory, California Institute of Technology, 
under contract with the National Aeronautics and Space Administration. 
This research also made use of APLpy, an open-source plotting package for Python 
hosted at http://aplpy.github.com, NASA's Astrophysics Data System and
IRAF, which is distributed by the National Optical Astronomy Observatory and 
operated by the Association of Universities for Research in Astronomy (AURA) under cooperative 
agreement with the National Science Foundation. 
The IAC researchers acknowledge financial support from the Spanish Ministry of Economy and Competitiveness 
(MINECO) under the 2011 Severo Ochoa Program MINECO SEV-2011-0187. CG and AA are partially funded by 
the Science and technology Ministry of Spain (grant AYA 2010-16717).
MB acknowledges support from grant AYA2013-48226-C3-1-P from the Spanish Ministry of Economy 
and Competitiveness (MINECO).



\end{document}